\documentclass[twoside,twocolumn,9pt]{article}
\usepackage{extsizes}
\usepackage[super,sort&compress,comma]{natbib} 
\usepackage[version=3]{mhchem}
\usepackage[left=1.5cm, right=1.5cm, top=1.785cm, bottom=2.0cm]{geometry}
\usepackage{balance}
\usepackage{widetext}
\usepackage{times,mathptmx}
\usepackage{sectsty}
\usepackage{graphicx} 
\usepackage{lastpage}
\usepackage[format=plain,font={stretch=1.125,small,sf},labelfont=bf,labelsep=space]{caption}
\usepackage[font=footnotesize,textfont=footnotesize]{subcaption}
\usepackage{float}
\usepackage{fancyhdr}
\usepackage{fnpos}
\usepackage[english]{babel}
\usepackage{array}
\usepackage{droidsans}
\usepackage{charter}
\usepackage[T1]{fontenc}
\usepackage[usenames,dvipsnames]{xcolor}
\usepackage{setspace}
\usepackage[compact]{titlesec}
\usepackage{color}
\usepackage{soul}
\usepackage[normalem]{ulem}
\usepackage{amsmath,amssymb}
\usepackage{epstopdf}
\usepackage{url}

\definecolor{cream}{RGB}{222,217,201}

\graphicspath{{./img/}}
\DeclareGraphicsExtensions{.pdf,.jpeg,.png,.jpg,.tif}
\begin{document}

\pagestyle{fancy}
\thispagestyle{plain}
\fancypagestyle{plain}{

\fancyhead[C]{\includegraphics[width=18.5cm]{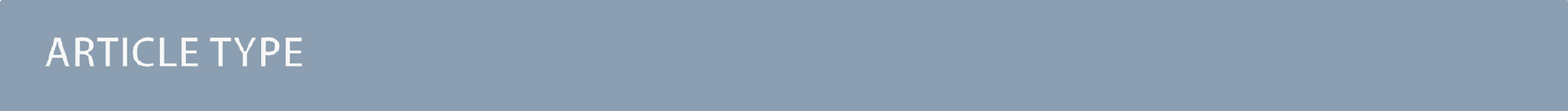}}
\fancyhead[L]{\hspace{0cm}\vspace{1.5cm}\includegraphics[height=30pt]{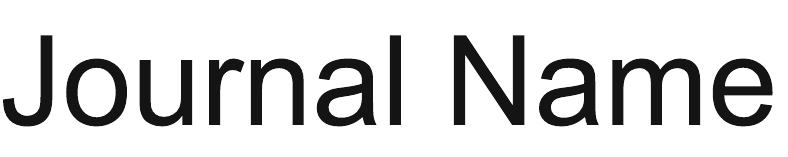}}
\fancyhead[R]{\hspace{0cm}\vspace{1.7cm}\includegraphics[height=55pt]{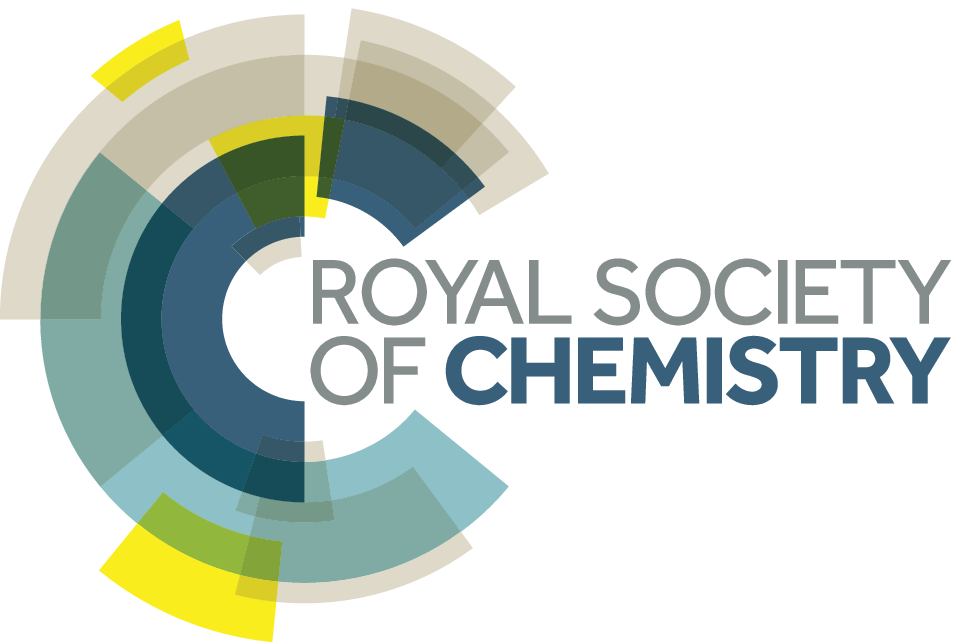}}
\renewcommand{\headrulewidth}{0pt}
}

\makeFNbottom
\makeatletter
\renewcommand\LARGE{\@setfontsize\LARGE{15pt}{17}}
\renewcommand\Large{\@setfontsize\Large{12pt}{14}}
\renewcommand\large{\@setfontsize\large{10pt}{12}}
\renewcommand\footnotesize{\@setfontsize\footnotesize{7pt}{10}}
\makeatother

\renewcommand{\thefootnote}{\fnsymbol{footnote}}
\renewcommand\footnoterule{\vspace*{1pt}%
\color{cream}\hrule width 3.5in height 0.4pt \color{black}\vspace*{5pt}} 
\setcounter{secnumdepth}{5}

\makeatletter 
\renewcommand\@biblabel[1]{#1}            
\renewcommand\@makefntext[1]%
{\noindent\makebox[0pt][r]{\@thefnmark\,}#1}
\makeatother 
\renewcommand{\figurename}{\small{Fig.}~}
\sectionfont{\sffamily\Large}
\subsectionfont{\normalsize}
\subsubsectionfont{\bf}
\setstretch{1.125} 
\setlength{\skip\footins}{0.8cm}
\setlength{\footnotesep}{0.25cm}
\setlength{\jot}{10pt}
\titlespacing*{\section}{0pt}{4pt}{4pt}
\titlespacing*{\subsection}{0pt}{15pt}{1pt}

\fancyfoot{}
\fancyfoot[LO,RE]{\vspace{-7.1pt}\includegraphics[height=9pt]{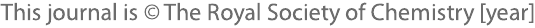}}
\fancyfoot[CO]{\vspace{-7.1pt}\hspace{13.2cm}\includegraphics{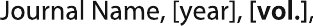}}
\fancyfoot[CE]{\vspace{-7.2pt}\hspace{-14.2cm}\includegraphics{head_foot/RF}}
\fancyfoot[RO]{\footnotesize{\sffamily{1--\pageref{LastPage} ~\textbar  \hspace{2pt}\thepage}}}
\fancyfoot[LE]{\footnotesize{\sffamily{\thepage~\textbar\hspace{3.45cm} 1--\pageref{LastPage}}}}
\fancyhead{}
\renewcommand{\headrulewidth}{0pt} 
\renewcommand{\footrulewidth}{0pt}
\setlength{\arrayrulewidth}{1pt}
\setlength{\columnsep}{6.5mm}
\setlength\bibsep{1pt}

\makeatletter 
\newlength{\figrulesep} 
\setlength{\figrulesep}{0.5\textfloatsep} 

\newcommand{\topfigrule}{\vspace*{-1pt}%
\noindent{\color{cream}\rule[-\figrulesep]{\columnwidth}{1.5pt}} }

\newcommand{\botfigrule}{\vspace*{-2pt}%
\noindent{\color{cream}\rule[\figrulesep]{\columnwidth}{1.5pt}} }

\newcommand{\dblfigrule}{\vspace*{-1pt}%
\noindent{\color{cream}\rule[-\figrulesep]{\textwidth}{1.5pt}} }

\makeatother


\twocolumn[
\begin{@twocolumnfalse}
  \vspace{3cm}
  \sffamily
  \begin{tabular}{m{4.5cm} p{13.5cm}}

    \includegraphics{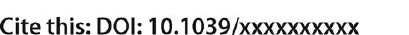} & \noindent\LARGE{\textbf{A neuromorphic systems approach to in-memory computing with non-ideal memristive devices: \newline From mitigation to exploitation}$^\dag$} \\
    \vspace{0.3cm} & \vspace{0.3cm} \\
                                    & \noindent\large{Melika Payvand,$^{\ast}$\textit{$^{a}$} Manu V Nair,\textit{$^{a\ddag}$} Lorenz K. Muller,$^{\ast}$\textit{$^{a}$}and Giacomo Indiveri\textit{$^{a}$}} \\

    \includegraphics{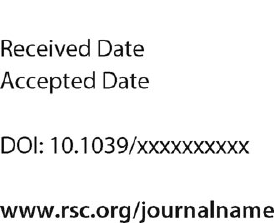} & \noindent\normalsize{Memristive devices represent a promising technology for building neuromorphic electronic systems. In addition to their compactness and non-volatility features, they are characterized by computationally relevant physical properties, such as state-dependence, non-linear conductance changes, and intrinsic variability in both their switching threshold and conductance values, that make them ideal devices for emulating the bio-physics of real synapses. In this paper we present a spiking neural network architecture that supports the use of memristive devices as synaptic elements, and propose mixed-signal analog-digital interfacing circuits which mitigate the effect of variability in their conductance values and exploit their variability in the switching threshold, for implementing stochastic learning. The effect of device variability is mitigated by using pairs of memristive devices configured in a complementary push-pull mechanism and interfaced to a current-mode normalizer circuit. The stochastic learning mechanism is obtained by mapping the desired change in synaptic weight into a corresponding switching probability that is derived from the intrinsic stochastic behavior of memristive devices. We demonstrate the features of the CMOS circuits and apply the architecture proposed to a standard neural network hand-written digit classification benchmark based on the MNIST data-set. We evaluate the performance of the approach proposed on this benchmark using behavioral-level spiking neural network simulation, showing both the effect of the reduction in conductance variability produced by the current-mode normalizer circuit, and the increase in performance as a function of the number of memristive devices used in each synapse.}
  \end{tabular}
\end{@twocolumnfalse} \vspace{0.6cm}
]

\renewcommand*\rmdefault{bch}\normalfont\upshape
\rmfamily
\section*{}
\vspace{-1cm}

\footnotetext{\textit{$^{a}$~Address, Address, Town, Country. Fax: XX XXXX XXXX; Tel: XX XXXX XXXX; E-mail: xxxx@aaa.bbb.ccc}}
\footnotetext{\textit{$^{b}$~Address, Address, Town, Country. }}

\footnotetext{\dag~Electronic Supplementary Information (ESI) available: [details of any supplementary information available should be included here]. See DOI: 10.1039/b000000x/}

\footnotetext{\ddag~Additional footnotes to the title and authors can be included \emph{e.g.}\ `Present address:' or `These authors contributed equally to this work' as above using the symbols: \ddag, \textsection, and \P. Please place the appropriate symbol next tibo the author's name and include a \texttt{\textbackslash footnotetext} entry in the the correct place in the list.}


Neuromorphic computing systems comprise synapse and neuron circuits arranged in a massively parallel manner to support the emulation of large-scale spiking neural networks~\cite{Chicca_etal14,Park_etal14,Furber_etal14,Benjamin_etal14,Merolla_etal14,Mitra_etal09,Qiao_etal15,Moradi_etal17,Davies_etal18}. In many of these systems, and in particular in neuromorphic processing devices designed to overcome the von-Neumann bottleneck problem~\cite{Backus78,Indiveri_Liu15,Qiao_etal15,Moradi_etal17,Boybat_etal18,Li_etal18,Ambrogio_etal18}, the bulk of the silicon real-estate is taken up by synaptic circuits that integrate in the same area both memory and computational primitives. To save area and maximize density in such devices, one possible approach is to implement very basic synapse circuits arranged in dense cross-bar arrays~\cite{Likharev_etal03,Linn_etal10,Kim_etal12,Prezioso_etal15,Sandrini_etal16}. However, such approach is likely to relegate the role of the synapse to a basic multiplier~\cite{Merolla_etal14a,Ambrogio_etal18}. In biology, synapses are extremely sophisticated structures that exhibit complex and powerful computational properties, including temporal dynamics, state-dependence, and stochastic learning behavior. The challenge is to design neuromorphic circuits that emulate these computational properties, and are also compact and low power. Memristive devices have recently emerged as nano-scale devices which provide a promising technology for addressing these problems~\cite{Yang_etal13,Indiveri_etal13}. These devices offer a compact and efficient solution to model synaptic weights since they are non-volatile, have a nano-scale footprint, can be integrated with Complementary Metal-Oxide Semiconductor (CMOS) chips~\cite{Payvand_etal15,Chakrabarti_etal17}, might only require little energy to change their state~\cite{Kim_etal12}, and in addition can emulate many of the synaptic functions observed in biological synapses~\cite{Jo_etal10,Prezioso_etal15,Kim_etal12}.
However, these devices are also characterized by non-idealities that introduce significant challenges in designing neural network architectures applied to classification and recognition tasks.
In particular, one property of memristive devices that introduces significant challenges in the design of large scale neural network architectures is the large variability of their operational parameters.
Memristive device variability exhibits itself in different forms, both between \emph{device to device} (spatial) and from \emph{cycle to cycle} within a single device (temporal). This variability therefore manifests itself both in the device conductance values and in their switching voltage~\cite{Ielmini_Waser15}. Device-to-device variability originates from process variations which also exists in current CMOS process, while the cycle-to-cycle variability stems from the underlying switching mechanism of memristors. 
The cycle-to-cycle variability is observed in different types of memristors, from Phase Change Memories (PCMs) ~\cite{Tuma_etal16} and Conductive Bridge RAMs ~\cite{Suri_etal13,Suri_etal12}to ionic redox-based resistive RAMs ~\cite{Gaba_etal13}. In particular, in the latter case, the underlying mechanism for this variability is associated with the formation and rupture of a conducting filament. Filament formation involves oxidation, ion transport and reduction which are all thermodynamical processes and as a result require overcoming an energy barrier. Therefore, the switching involves thermal activation to surpass the barrier and thus is a probabilistic process. In other words, for the same devices and the same filament, the nature of the switching events will occur randomly and is thus \emph{stochastic}.~\cite{Jo_etal08,Gaba_etal13,Ambrogio_etal16a,Yang_etal13}.

To summarize, the variability in memristive devices results in a distribution of different parameters that can be categorized in four distinct groups:
\begin{description}
\item[G1] Distribution of the switching voltage of a single device
\item[G2] Distribution of the high and low resistive states of a single device 
\item[G3] Distribution of the switching voltages among multiple devices
\item[G4] Distribution of the high and low resistive states among multiple devices
\end{description}

The variability of parameters across multiple devices (e.g., for groups G3 and G4) can be mitigated and managed for example by considering only binary states~\cite{Truong_etal14}, by implementing ``compound'' synapses that employ multiple memristive devices per synaptic element~\cite{Serb_etal16,Boybat_etal18}, or by interfacing the memristive devices to CMOS processing stages that reduce the effect of their variability~\cite{Nair_Indiveri17}. Conversely, the cycle-to-cycle variability (e.g., in groups G1 and G2) can be managed by using feedback control to set the desired state to a well-defined value~\cite{Serb_etal16b}, which requires a large overhead control circuit, or it can be exploited as a means to implement stochastic learning in spiking neural networks~\cite{Suri_etal13,Vincent_etal14,Al-Shedivat_etal15a,Tuma_etal16,Neftci_etal16,Payvand_etal18}. Indeed, it has been shown that employing binary synapses,  variability and randomness in their switching threshold in spiking neural networks greatly improves the convergence of the network and provides a form of regularization which substantially improves the network generalization performance~\cite{Suri_etal13,Bill_Legenstein14,Courbariaux_etal15}. In the case of neural networks with low resolution synapses, it has been shown that a randomized gradient descent method significantly outperforms naive deterministic rounding methods~\cite{Muller_Indiveri15}. 

Memristive devices are a promising emerging technology for use in large-scale neural network architectures~\cite{Boybat_etal18,Ambrogio_etal18,Wozniak_etal17,Covi_etal16,Serrano-Gotarredona_Linares-Barranco14,Indiveri_etal13}. 
Employing such devices in neural processing systems for robust computation in real-world practical applications calls for ways to either mitigate their non-idealities, to exploit them, or to combine the best of both approaches in the same architecture.
In this paper we present a spiking neural network architecture that support the use of variable and stochastic memristive devices for robust inference and probabilistic learning. We show that by combining such devices with state-of-the-art mixed-signal digital and analog subthreshold circuits~\cite{Liu_etal02a}, it is possible to build electronic learning systems with biologically plausible functionality which can process and classify sensory data directly on-chip in real-time, and which represent ideal technologies for always-on edge-computing neural network applications.
We propose synapse-CMOS interfacing circuits that dramatically reduce the effect of device-to-device variability, as well as spike-based learning circuits that are compatible and exploit the device cycle-to-cycle variability to implement stochastic learning. We validate the functionality of such circuits by applying the neural network architecture to a pattern classification task, using a standard digit recognition benchmark based on the Modified National Institute of Standards and Technology (MNIST) data-set~\cite{mnist}.
In the next section we describe the spiking neural network architecture, explain its basic principle of event-based operation,  and present its main neuromorphic building blocks; in Section~\ref{sec:mism-norm} we present the memristive synapse circuits and their related current-mode sense circuits used to reduce the device-to-device variability for improving the network performance in its inference phase; in Section~\ref{sec:stoch} we present the spike-based stochastic learning circuits that exploit the devices cycle-to-cycle variability for inducing probabilistic state changes in the network synaptic weights; Section~\ref{sec:sims} presents behavioral simulations results at the system level, in the hand-written digit recognition benchmark to validate the proposed circuits and approach; finally in Section~\ref{sec:conclusions} we present the concluding remarks.

\section{The neuromorphic architecture}
\label{sec:arch}

\begin{figure*}
 \centering
 \includegraphics[width=\textwidth]{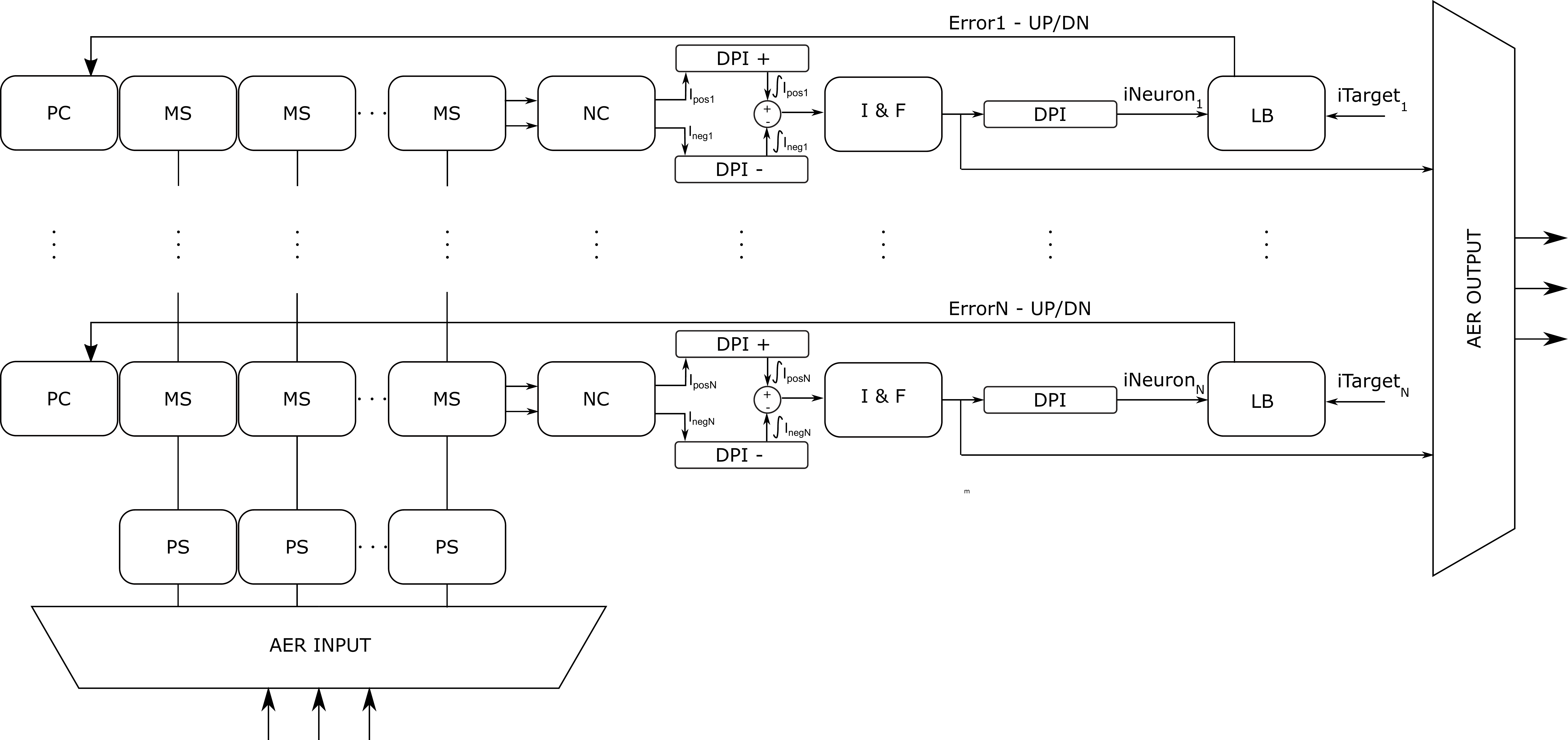}
 \caption{Neuromorphic architecture comprising multiple silicon neurons, each receiving inputs from CMOS-memristive synapse elements. MS is short for Memristive Synapse, PS for Pulse Shaper, NC for Normalizer Circuit, DPI for Differential Pair Integrator, I\&F Neuron for Integrate and Fire Neuron, LB for Learning Block and PC for Programming Circuitry. }
 \label{fig:arch}
\end{figure*}

\begin{figure}
 \centering
 \includegraphics[width=\linewidth]{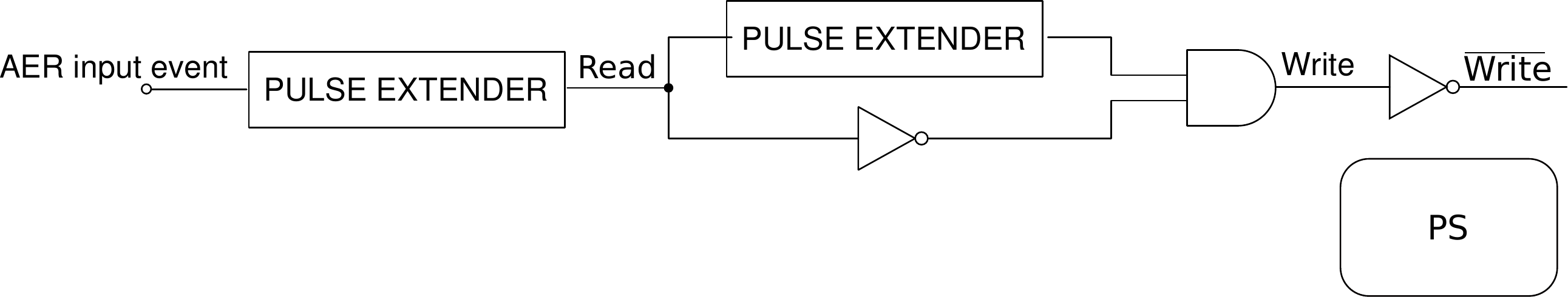}
 \caption{Pulse shaper (\textsf{PS}) block schematic. With the arrival of an input event from the \textsf{AER} block, two consecutive pulses \textsf{Read} and \textsf{Write} are generated by two digital Pulse Extender circuits.}
 \label{fig:pulse_shaper}
\end{figure}

\begin{figure}
 \centering
 \includegraphics[width=0.9\linewidth]{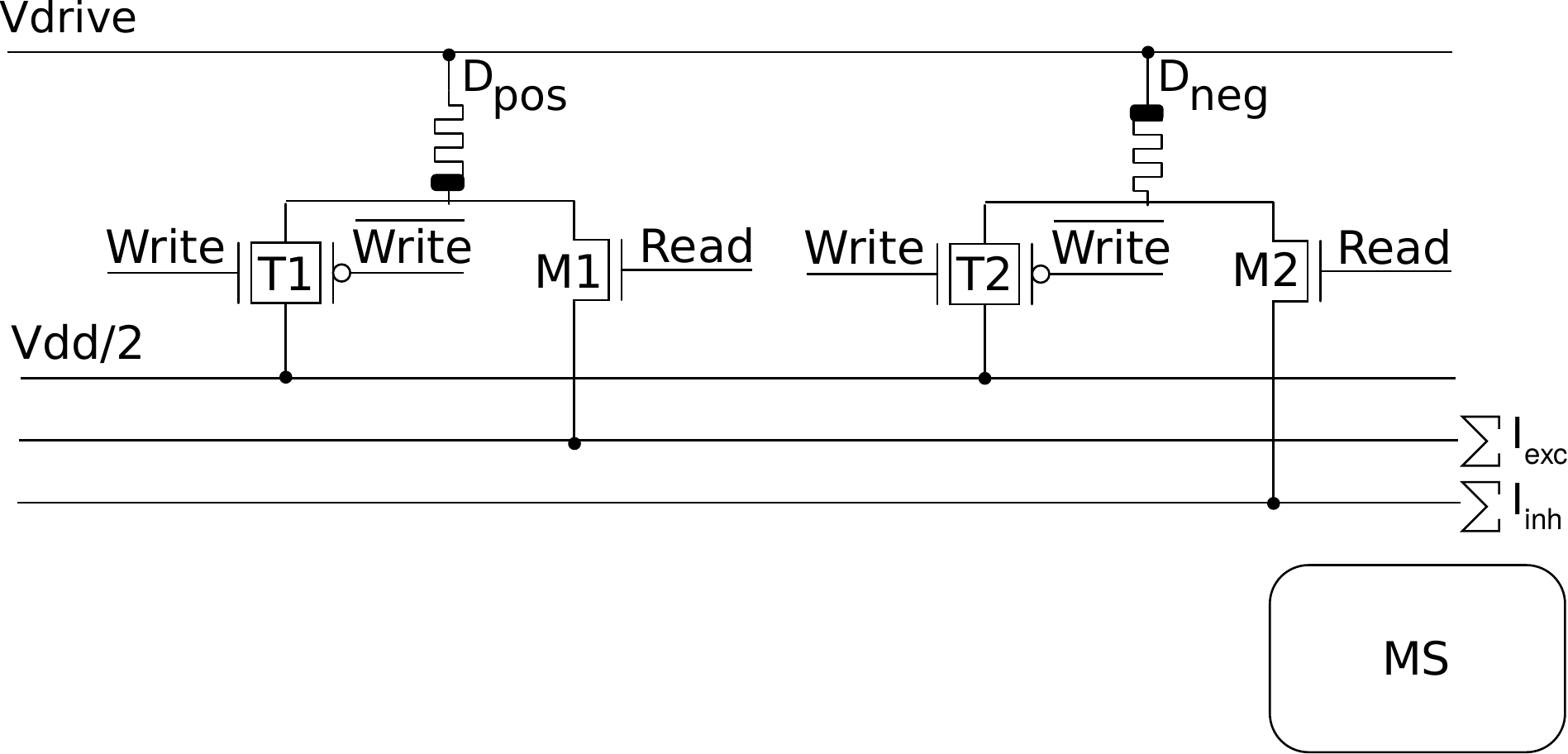}
 \caption{A single Memristive Synapse (\textsf{MS}) block of the proposed neuromorphic system. The devices $\mathsf{D_{pos}}$ and $\mathsf{D_{neg}}$ are modeling the excitatory and inhibitory synapses respectively. When the \textsf{Read} pulse signal from the corresponding column is active, the excitatory currents sum together on the excitatory $\mathsf{\sum I_{exc}}$ and the inhibitory $\mathsf{\sum I_{inh}}$ lines. Similarly, when the \textsf{Write} pulse is high, the switches connect the devices to the programming lines.}
 \label{fig:synapse}
\end{figure}

The spiking neural network architecture that supports the use of memristive circuits as synapse elements is shown in Fig.~\ref{fig:arch}. This architecture expects input spikes and produces output spikes that are encoded as Address-Events: each neuron is assigned a unique address, and when it produces an output spike, a corresponding digital pulse is encoded on a common shared time-multiplexed bus with its corresponding address. Potential collisions arising from multiple neurons requesting access to the same bus are handled by asynchronous arbiter circuits, that are part of the Address-Event Representation (AER) protocol~\cite{Deiss_etal98,Lazzaro_Wawrzynek95}. In this protocol, the analog information present in the silicon neuron is encoded in the time interval between its address-events. The asynchronous nature of this communication protocol ensures that precise timing information is preserved, and signals are transmitted only when there is neural activity. As neural activity in spiking neural networks is typically sparse in both space and time, this protocol is ideal for minimizing power-consumption and maximizing bandwidth~\cite{Boahen98}. The architecture of Fig.~\ref{fig:arch} comprises multiple rows of neurons, each composed of multiple Memristive Synapse (MR) elements, Integrate and Fire (I\&F) soma circuits, and additional interfacing circuits for managing the input pulse shapes, the synaptic currents, their temporal dynamics, and the spike-based learning mechanism. Upon the arrival of an input Address-Event, this is decoded by the AER input circuits into a one-hot pulse to be transmitted to the target column in the network. This decoded pulse is then converted by a dedicated Pulse Shaper (PS) circuit, which produces a \textsf{Read} and a \textsf{Write} pulse, used to measure the currents through the memristive synapse elements and potentially change their conductance values correspondingly. A schematic diagram of the PS circuit is shown in Fig.~\ref{fig:pulse_shaper}. The pulse extender circuit block in the figure is based on a classical a starved-inverter circuit, and has been characterized in previous work~\cite{Nair_etal17}.
The output of the PS block is then broadcast to all \textsf{MS} synapse blocks of the corresponding column. Each \textsf{MS} synapse comprises one pair of memristive devices arranged in a complementary configuration (see $\mathsf{D_{pos}}$ and $\mathsf{D_{neg}}$ of Fig.~\ref{fig:synapse}). The pairs of devices are arranged in a way to produce positive contributing currents (modeling  excitatory synapses) and negative contributing ones (modeling inhibitory synapses) during the ``read-phase'', and are updated in a push-pull way during the ``write-phase'' (i.e., if the conductance of one device is increased, the conductance of the complementary device is decreased, and vice-versa). Specifically, during the read phase, the $\mathsf{V_{drive}}$ voltage of Fig. \ref{fig:synapse} is set to a small value, such that small currents (e.g., of the order of nano-Ampere) will flow through the memristive pair onto the separate positive and negative summing lines. Conversely, during the write phase, digital control signals disable the connection to the current summing lines and enable the connection to the weight update Programming Circuits (\textsf{PC}), which set the $\mathsf{V_{drive}}$ signal to either $\mathsf{V_{dd}}$ or $\mathsf{Gnd}$ depending on the sign of \textsf{Error} signal produced by spike based learning Block (\textsf{LB}) of the corresponding row.

During the read phase, the output currents produced by all MS blocks along a row in the architecture are summed through Kirchhoff's current law and conveyed to a Normalizer Circuit (\textsf{NC}) block. This is a current-mode circuit based on the Gilbert normalizer circuit~\cite{Gilbert96} which receives the positive and negative contributions of currents from the memristive devices and produces two corresponding output currents that are scaled and normalized appropriately. As this circuit plays a fundamental role in reducing the effect of device variability across all memristive devices present in the neuron row, we describe its functionality in detail in Section~\ref{sec:mism-norm}.

The positive and negative output currents produced by the NC block are then sent to two separate Differential Pair Integrator (\textsf{DPI}) circuits~\cite{Bartolozzi_Indiveri07a}. These are current-mode linear integrator filters that integrate the incoming current pulses and produce temporally decaying currents that faithfully model the Excitatory Post Synaptic Current (EPSC) and Inhibitory Post Synaptic Current (IPSC)  counterparts of real biological synapses. The difference between positive and negative synaptic current contributions is then sent into the \textsf{I\&F} soma block, that temporally integrates these currents and produces an output spike as soon as the integrated current reaches the neuron's firing threshold.  Both DPI and I\&F blocks have been fully characterized and explained in a previous work~\cite{Chicca_etal14}.

The output spikes of the \textsf{I\&F} block are sent to the AER output circuits, as well as to an additional DPI circuit that integrates the neurons spikes. The output current of this DPI circuit (see \textsf{iNeuron} of Fig.~\ref{fig:arch}) is proportional to the neuron's average firing rate.  It is sent as input to the neuron's Learning Block (\textsf{LB}), which compares the neuron's output firing rate to a desired target value, and produces an error signal that is proportional to the difference. This error signal is then used by the corresponding row Programming Circuit (\textsf{PC}) block to change the probability of synaptic weight update in the synapses that were stimulated by the incoming Address-Event. These circuits implement the  probabilistic ``Delta'' learning rule~\cite{Widrow_Hoff60} used in the architecture, and they are fully described in Section~\ref{sec:stoch}.

\section{The memristive current normalizer circuit}
\label{sec:mism-norm}

\begin{figure}
  \centering
  \includegraphics[width=0.45\linewidth]{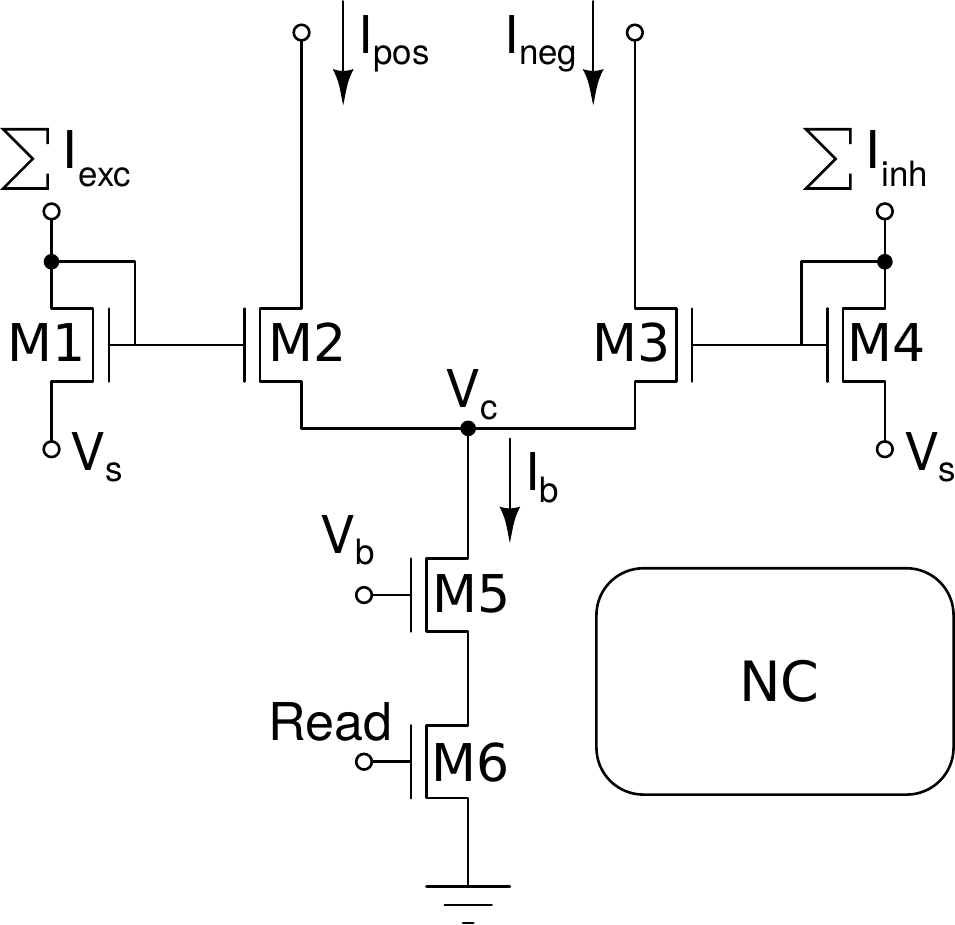}
  \caption{Current-mode normalizer circuit (\textsf{NC}) block. Input currents coming from multiple synapses from the excitatory and inhibitory lines are scaled and normalized.}
  \label{fig:mem_cell}
\end{figure}

\begin{figure}
  \begin{subfigure}{0.45\textwidth}
    \centering
    \includegraphics[width=\textwidth]{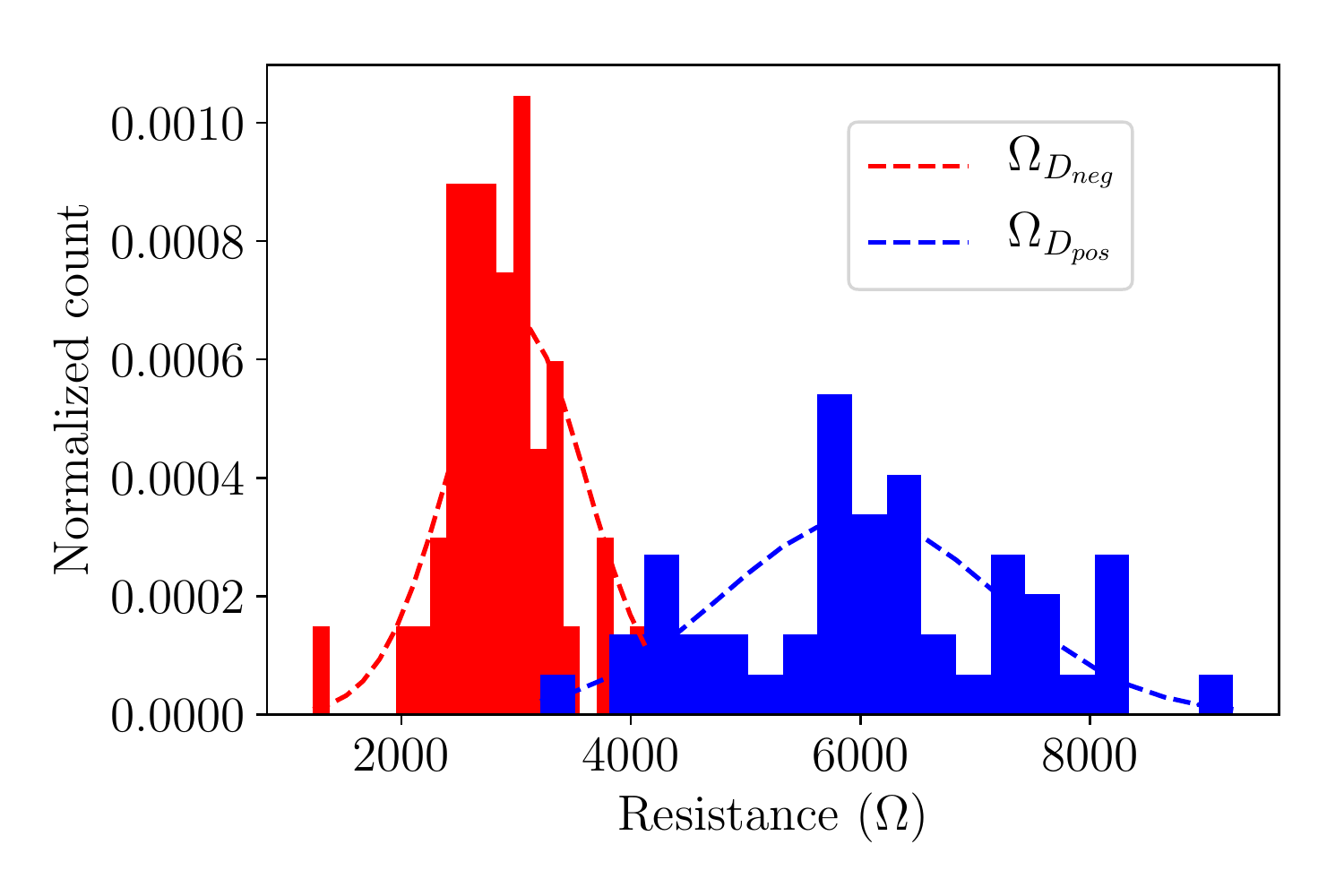}
    \subcaption{}
    \label{fig:mismatch_res_var}
  \end{subfigure}
  \begin{subfigure}{0.45\textwidth}
    \centering
    \includegraphics[width=\textwidth]{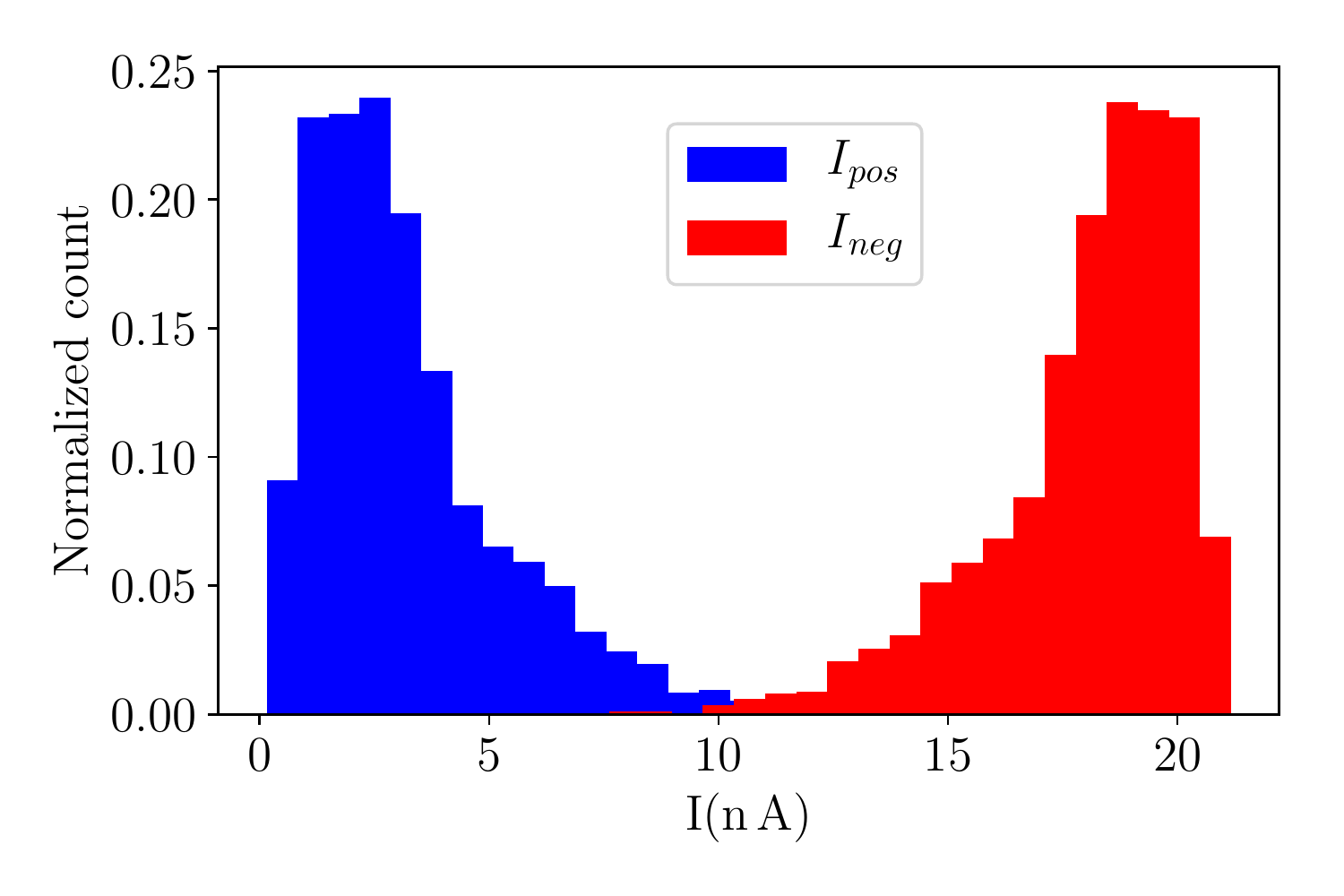}
    \subcaption{}
    \label{fig:tf}
  \end{subfigure}
   \caption{Histograms highlighting the differential memristive synapse
    weight storage behavior for on/off resistance ratio of 2: (2.87$k\Omega$, 490$\Omega$),
    $\Omega_{D_{pos}}$ = (6.12$k\Omega$, 1.3$k\Omega$). Monte Carlo circuit simulations were run to obtain these plots where 50 values of low and high conductance states were sampled and plotted in 20 bins. Dashed lines show the sampling distributions for device high and low conductance states in (\subref{fig:mismatch_res_var}). (\subref{fig:tf}) shows the distribution of the output currents from the normalizer circuit. The shown histograms are normalized by dividing the count by the number of observations times the bin width.}
  \label{fig:2xcase}
\end{figure}
The memristive current normalizer circuit is shown in Fig.~\ref{fig:mem_cell}. The circuit is operated in the weak inversion, or subthreshold domain~\cite{Liu_etal02a} where transistors have an exponential transfer function, in order to reproduce the functionality of the Gilbert-normalizer element~\cite{Gilbert90} which was originally designed for use with bi-polar transistors.
The input signals to this circuit are given by the sum of the currents measured across the memristive devices in the corresponding neuron row (see also Fig.~\ref{fig:arch}). The circuit has a differential input, provided by the positive and negative summing lines of the circuit's row. As these input currents are proportional to the values of the memristive devices, they can be affected by a large variation in their values. However, it has been demonstrated~\cite{Nair_etal17} that the normalizer output currents $I_{pos}$ and $I_{neg}$ of Fig.~\ref{fig:mem_cell}, can be approximately expressed as function of the input currents $\sum I_{exc}$ and $\sum I_{inh}$, which in turn are proportional to the  memristive device conductances:
\begin{align}
  I_{pos} & = I_b\frac{\sum I_{exc}}{\sum I_{exc} + \sum I_{inh}} &   I_{neg} & = I_b\frac{\sum I_{inh}}{\sum I_{exc} + \sum I_{inh}}
\label{eq:norm}   
\end{align}
Since in each Memristive Synapse block the memristive devices are arranged in a push-pull configuration (see Fig.~\ref{fig:synapse}), large $\sum I_{exc}$ currents will typically result in small $\sum I_{inh}$ currents and vice-versa. In the extreme case, when all conductances of one type (e.g., excitatory) are in the high state and the conductances of the other type (e.g., inhibitory) are in the low state, one output current of the circuit will be approximately equal to the maximum possible value (e.g., $I_{pos}\approx I_{b}$)  and the other to the minimum value, which is set by the transistor leakage current. It is due to this strong non-linear behavior that the normalizing function of eq.~\eqref{eq:norm} has the remarkable effect of reducing the effect of device mismatch in their conductance values. Examples of the variability reduction features of the circuit are illustrated in Figures~\ref{fig:2xcase} and~\ref{fig:10xcase}:
\begin{figure}
  \begin{subfigure}{0.45\textwidth}
    \centering \includegraphics[width=\textwidth]{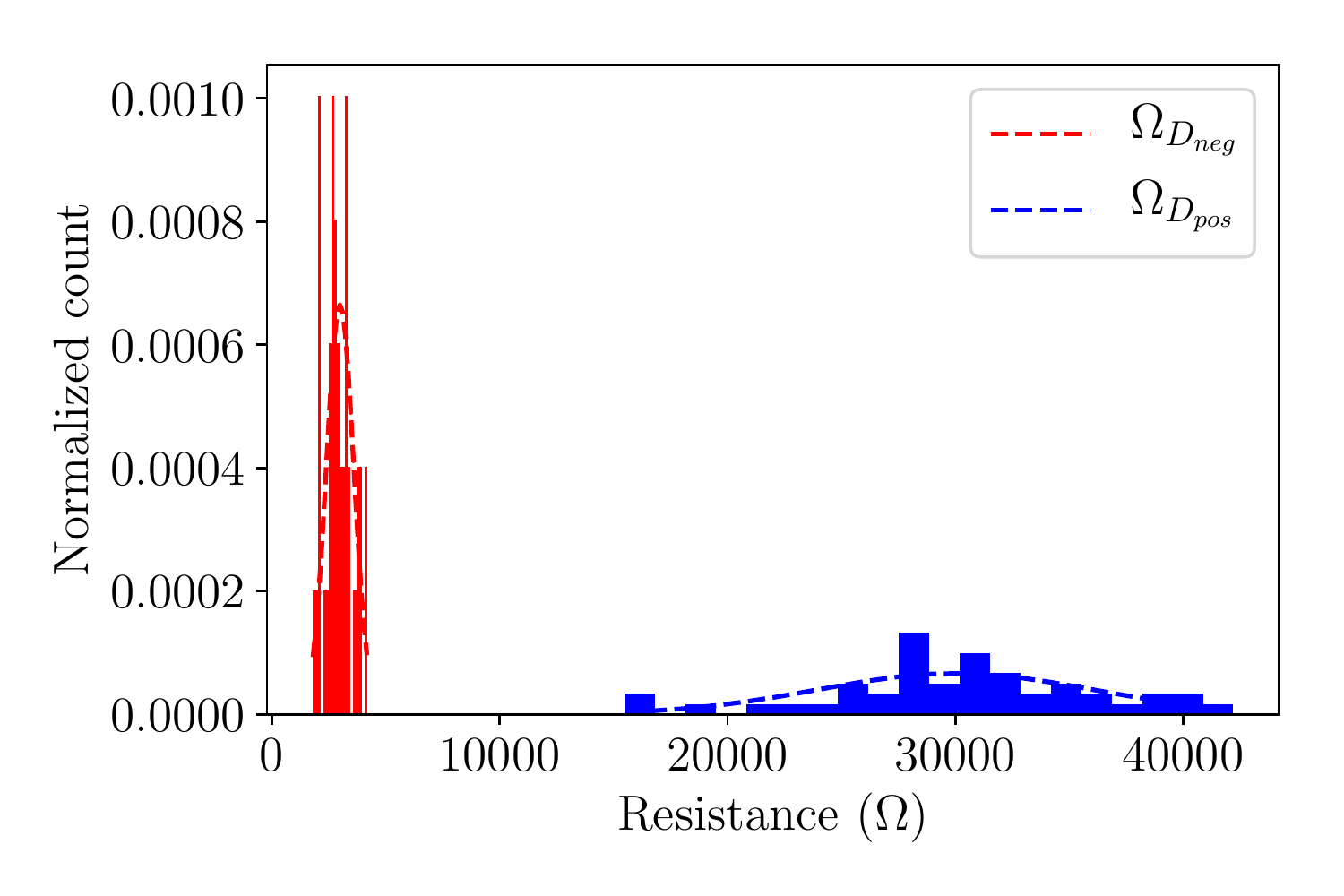}
    \subcaption{}
    \label{fig:mismatch_res_var10}
  \end{subfigure}
  \begin{subfigure}{0.45\textwidth}
    \centering \includegraphics[width=\textwidth]{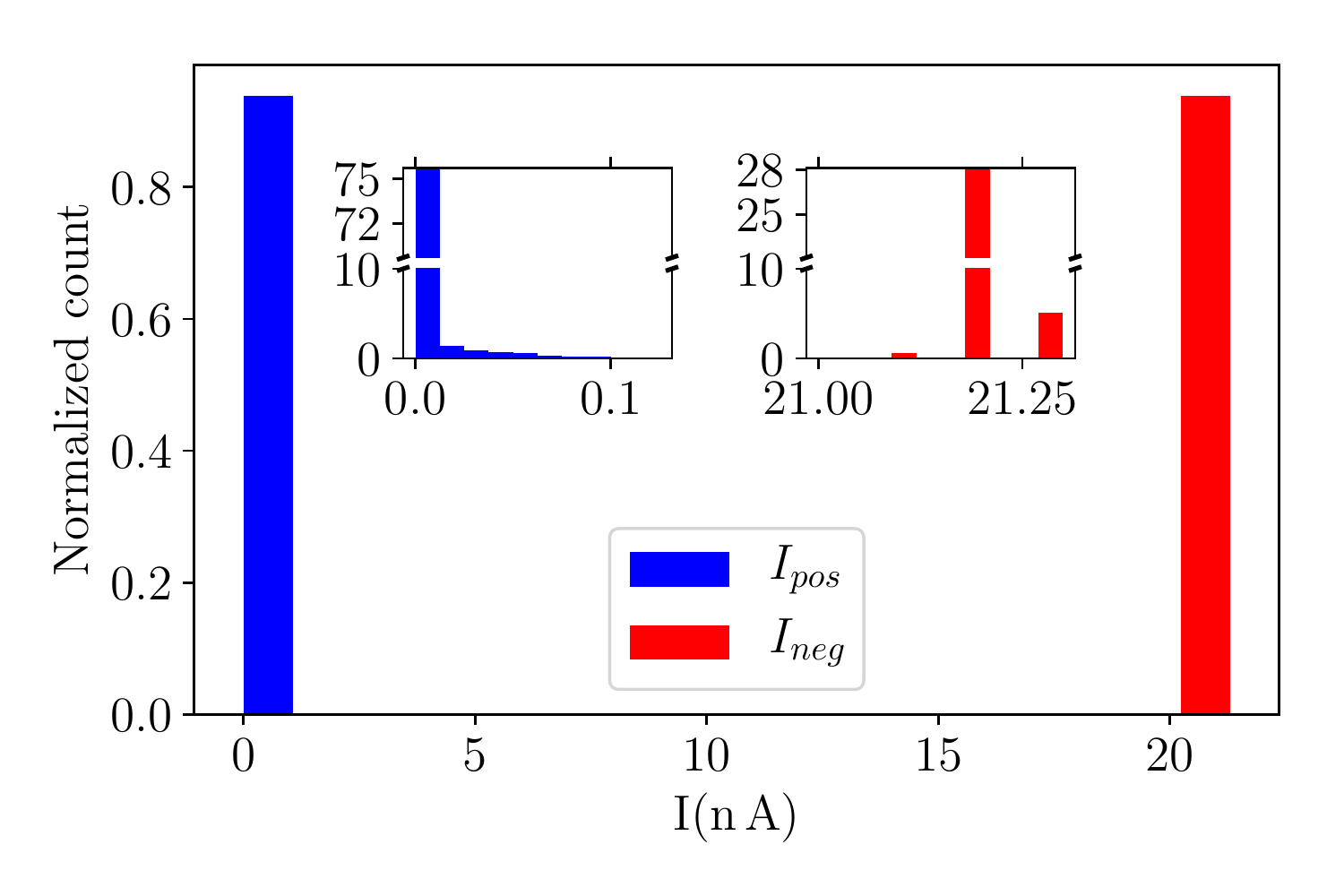}
    \subcaption{}
    \label{fig:tf10}
  \end{subfigure}
   \caption{Histograms highlighting the differential memristive synapse synaptic
    weight storage behavior for high/low resistance ratio of 10: (Mean, Std
    Dev) for $\Omega_{D_{neg}}$ = (2.931$k\Omega$, 582$\Omega$),
    $\Omega_{D_{pos}}$ = (30.35$k\Omega$, 5.71$k\Omega$). Monte Carlo circuit simulations were run to obtain these plots where 50 values of low and high conductance states were sampled and plotted in 20 bins. Dashed lines show the sampling distributions for device high and low conductance states in (\subref{fig:mismatch_res_var}). (\subref{fig:tf}) shows the distribution of the output currents from the normalizer circuit. The insets in Figure~\ref{fig:tf10} show the resulting output current distributions in finer detail where the range of observed values for $I_{pos}$ and $I_{neg}$ are plotted in 10 bins without normalization.}
     \label{fig:10xcase}
   \end{figure}
Figure~\ref{fig:mismatch_res_var} shows the effect of the normalizer circuit on its output currents for a typical distribution of device conductances that was derived from the literature~\cite{Brivio_etal16}, for a very conservative on-off ratio of two. While there is a significant overlap between the resistance values of the single memristive devices (see Fig.~\ref{fig:mismatch_res_var}), it is clear from Fig.~\ref{fig:tf} that using the output of the normalizer to measure synaptic weight values reduces this overlap significantly, as it squashes the distributions of output currents toward the maximum and minimum possible current outputs. This is even more evident in Fig.~\ref{fig:10xcase}, where the on-off ratio of the conductance values is ten. In particular, note that in this case the normalizer circuit eliminates the effect of device variability almost completely, as the distributions of currents (equivalent to the distribution of synaptic weights) is almost completely binary, despite the fact that the distribution in memristive conductance values is still substantial (compare Fig.~\ref{fig:mismatch_res_var10} with Fig.~\ref{fig:tf10}).

As the output currents of the normalizer circuit can be scaled to very small subthreshold current values (e.g., in the range of pico-Amperes), the power consumption of the neural processing circuits downstream can be kept very low. Furthermore, this makes the downstream circuits more compact as they can use smaller capacitors to implement temporal dynamics with biologically plausible time constants (e.g., for allowing real-time interaction with the environment). In addition to mitigating the effect of device variability, the differential operation used in the architecture proposed has the advantage of allowing the use of both positive (excitatory) and negative (inhibitory) weights, effectively doubling the ``high-low'' dynamic range of the memristive devices.

\section{The stochastic learning circuits}
\label{sec:stoch}

In this section we propose circuits that can be interfaced to memristive devices to exploit the cycle-to-cycle variability in their switching characteristics to implement stochastic learning. Indeed, the cycle-to-cycle variability in the switching of memristors provides an intrinsic stochastic process that can be used to update the weights of the synapses in a neural network.
The probabilistic switching in the memristor devices has been observed and studied before which is believed to stem from the formation and dissolution of a filament between the device electrodes~\cite{Jo_etal08, Gaba_etal13, Naous_etal16}. 
The  filament formation model in the memristive devices is strongly bias-dependent and can be explained by the hopping of ions in a thermally activated process~\cite{Jo_etal08}. The hopping rate is therefore exponentially related to the activation energy and linearly dependent in time:
\begin{equation}
  \Gamma=1/\tau=\upsilon e^{-E_a(V)/k_BT},
\end{equation}
where $\upsilon$ is the attempt frequency for particle hopping, $k_B$ is the Boltzmann constant and $T$ is the absolute temperature. As a result of the thermodynamical nature of this process, the switching of the memristive devices is stochastic and is shown to follow a Poisson distribution in silver/amorphous silicon/p doped poly silicon memristive devices~\cite{Gaba_etal13}. The authors claim that the results can be generalized to other memristive systems such as OxRAMs. The Poisson distribution suggests that the switching events are independent from one another and that the probability of a switching event occurring within $\Delta t$ at time t is  $P(t)=\frac{\Delta t}{\tau} e^{-t/\tau}$, where $\tau$ is the characteristic wait time which is the mean time after the application of the SET pulse in which the device switches. A thorough study on the effect of the applied SET voltage $V$ on the wait time has been  performed which shows that as the applied voltage across the device increases linearly, the characteristic wait time decreases exponentially~\cite{Gaba_etal13}.
Therefore, $\tau(V)=\tau_0 e^{-V/V_0}$ where $\tau_0$ and $V_0$ are fitting parameters found by the experimental measurements~\cite{Jo_etal08}.
Employing this model, the probability of switching for $t<<\tau$ can be written as~\cite{Naous_etal16}:
\begin{equation}
  P(t)=\frac{\Delta t}{\tau}=\frac{\Delta t}{\tau_0 e^{V/V_0}}
  \label{eq:pswitch}
\end{equation}

The stochastic learning mechanism we propose exploits this characteristic in an event-based network which comprises binary synapses, implemented using memristive devices that are driven to their maximum or minimum conductance states with every weight update. Even though the synapses are treated as binary elements, the probabilistic nature of the weight-update mechanism can be used to preserve the analog nature of the learning rule.
The weight update mechanism that we consider in this work is the ``Delta-rule''.
This is one of the most common weight update rules used in the literature for single-layer networks~\cite{Widrow_Hoff60,Hertz_etal91}, and it is at the base of the back-propagation algorithm used in the vast majority of current multi-layer neural networks~\cite{LeCun_etal15,Schmidhuber15}.
It has been shown that the Delta-rule is a learning algorithm which minimizes the Least Mean Square (LMS) error of a single-layer neural network cost function defined as the difference between a target desired output signal $T$ and the network output signal $y$,  for a given set of input patterns signals $x$, weighted by the synaptic weight parameters $w$. Specifically, this learning rule sets the corresponding weight change between the $ith$ input and the $jth$ output neuron to be: $\Delta w_{ji}=\alpha(T_j-y_j)x_i$~\cite{Hertz_etal91}. 

In the stochastic version of the Delta-rule, this weight update is translated to the \emph{probability of weight change}, and in the context of implementing it with memristive devices, to the \emph{probability of switching} the device's state rather than an incremental change in its conductance. Therefore, to directly map the probability $P$, into the weight change $\Delta w_{ji}$, $P$ has to be a linear function of the error $(T_j-y_j)$. Since from eq.~\eqref{eq:pswitch} , $P$ is an exponential function of the voltage applied across the device, this voltage needs to be:
\begin {equation} 
\frac{V}{V_0} = \log(T_j-y_j)
\label{eq:logV}.
\end{equation}

such that by plugging eq.~\eqref{eq:logV} into eq.~\eqref{eq:pswitch} we get:

\begin{eqnarray}
P(t) = \Delta t e^{\log(T_j-y_j)} x_i &= \Delta t (T_j-y_j) x_i
\label{eq:prob-delta}
\end{eqnarray}

which ensures that $P$ follows a linear function of the error.

\begin{figure}
 \centering
 \includegraphics[width=0.35\textwidth]{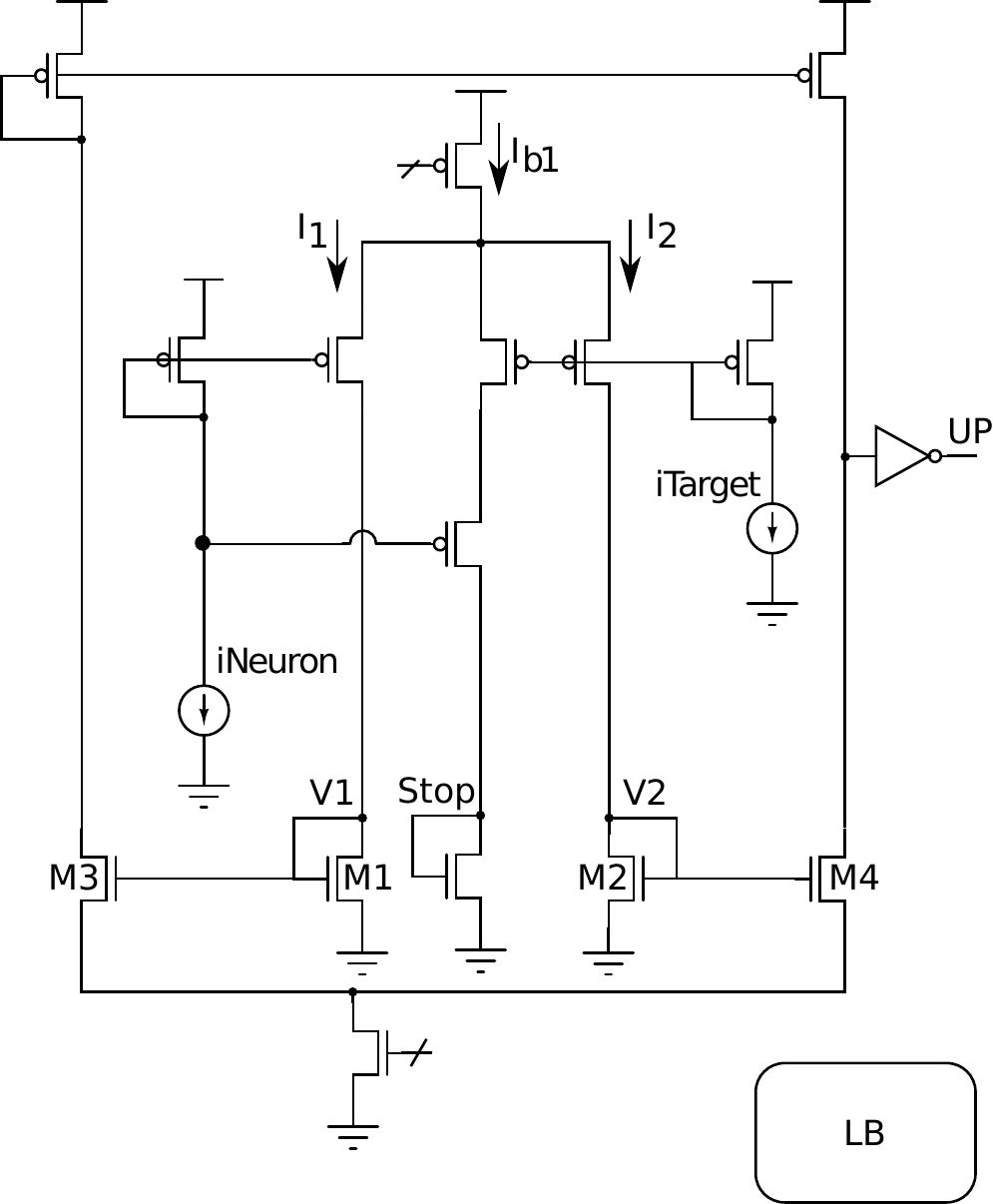}
 \caption{Learning Block circuit, implemented as a Bump/anti-Bump circuit. The neuron average activity \textsf{iNeuron} is compared against a target current \textsf{iTarget}. The voltages \textsf{V1} and \textsf{V2} are a function of the difference between \textsf{iTarget} and \textsf{iNeuron}. The digital signal \textsf{UP} is high when the error is positive and is low otherwise.}
 \label{fig:bump}
\end{figure}

In our framework we encode input signals $x$ as a sequence of pre-synaptic events coming from the AER block, which also trigger the weight update at their arrival.
The error signal used for the weight updates depends on the average firing rate of the output neuron (equivalent to the Delta-rule $y$ signal) and on a desired target signal  $T$ provided as an external input. The neuron average firing rate is computed using a current-mode low pass filter (see the DPI circuit of Fig.~\ref{fig:arch} which produces the current \textsf{iNeuron}). The desired target signal is represented by the current \textsf{iTarget}. 

To compute the error as the difference of these two signals, we used the circuit shown in Fig.~\ref{fig:bump}. 
It is an analog circuit operated in the subthreshold domain known as the  Bump/anti-Bump circuit~\cite{Liu_etal02a}. The circuit generates a current in the middle branch that increases as the values of \textsf{iNeuron} and \textsf{iTarget} become more and more similar (bump), whereas it generates increasing currents in the side branches as \textsf{iNeuron} and \textsf{iTarget} become dissimilar (anti-bump). Note that the side branch currents, labeled as $I1$ and $I2$ in Fig.~\ref{fig:bump}, have the same transfer function of the current normalizer circuit described in Section~\ref{sec:mism-norm}:

\begin{equation}
  I_1=I_{b1} \frac{iNeuron}{iNeuron+iTarget} ;
  \quad
  I_2=I_{b1} \frac{iTarget}{iNeuron+iTarget} 
\end{equation}

The difference in the side currents is then thresholded and digitized to produce a digital control signal \textsl{UP}, and its inverse \textsf{DN} (not shown in the figure), that controls the direction of the weight update for the synapse that received its corresponding input event.

The voltage applied to the memristive devices to implement the probabilistic weight change of eq.~\eqref{eq:prob-delta} is determined by eq.~\eqref{eq:logV}.
The precise value of this voltage is very important, as the probability of switching of a memristor is exponentially dependent on the voltage across it.
However CMOS device mismatch and memristive device variability do not allow the use of a single constant voltage shared across all synapses.
Although analogous efforts have been proposed in the literature~\cite{Schemmel_etal10}, implementing calibration circuits to precisely control the voltage biases in each synapse circuit would result in a very bulky design with large overhead circuitry and time-consuming calibration procedures at run time.

Rather than attempting to solve the device mismatch and variability effects with brute-force approaches, we exploit the stochastic nature of the learning algorithm:  by generating a time-varying voltage ramp signal and applying it to the memristive devices in the weight-update phase, we can sweep across all values of the distribution of voltages that can affect the device switching behavior. 
Specifically, we propose a circuit that generates a ramp voltage with a slope $\alpha$ that is proportional to the logarithmic value of the error signal, as defined in eq.\eqref{eq:logV}.

By applying this voltage ramp to the memristive devices, the switching probability of the devices becomes proportional to \textsf{iTarget-iNeuron}. Since \textsf{iTarget} is the desired output spike rate and \textsf{iNeuron} the effective output spike rate, the expected weight change resulting from a switching is thus proportional to the derivative of this difference squared: In expectation the circuit implements a gradient descent procedure on this squared error. 
The time varying ramp signal modulates the probability of resistive switching such that high errors results in more probable switching and vice versa.

This strategy implements a form of ``Randomized Rounding''~\cite{Raghavan_Tompson87} on the Delta-Rule, which has been shown to be more effective than deterministic rounding in a similar context~\cite{Muller_Indiveri15}.
\begin{figure}
 \centering
 \includegraphics[width=0.5\textwidth]{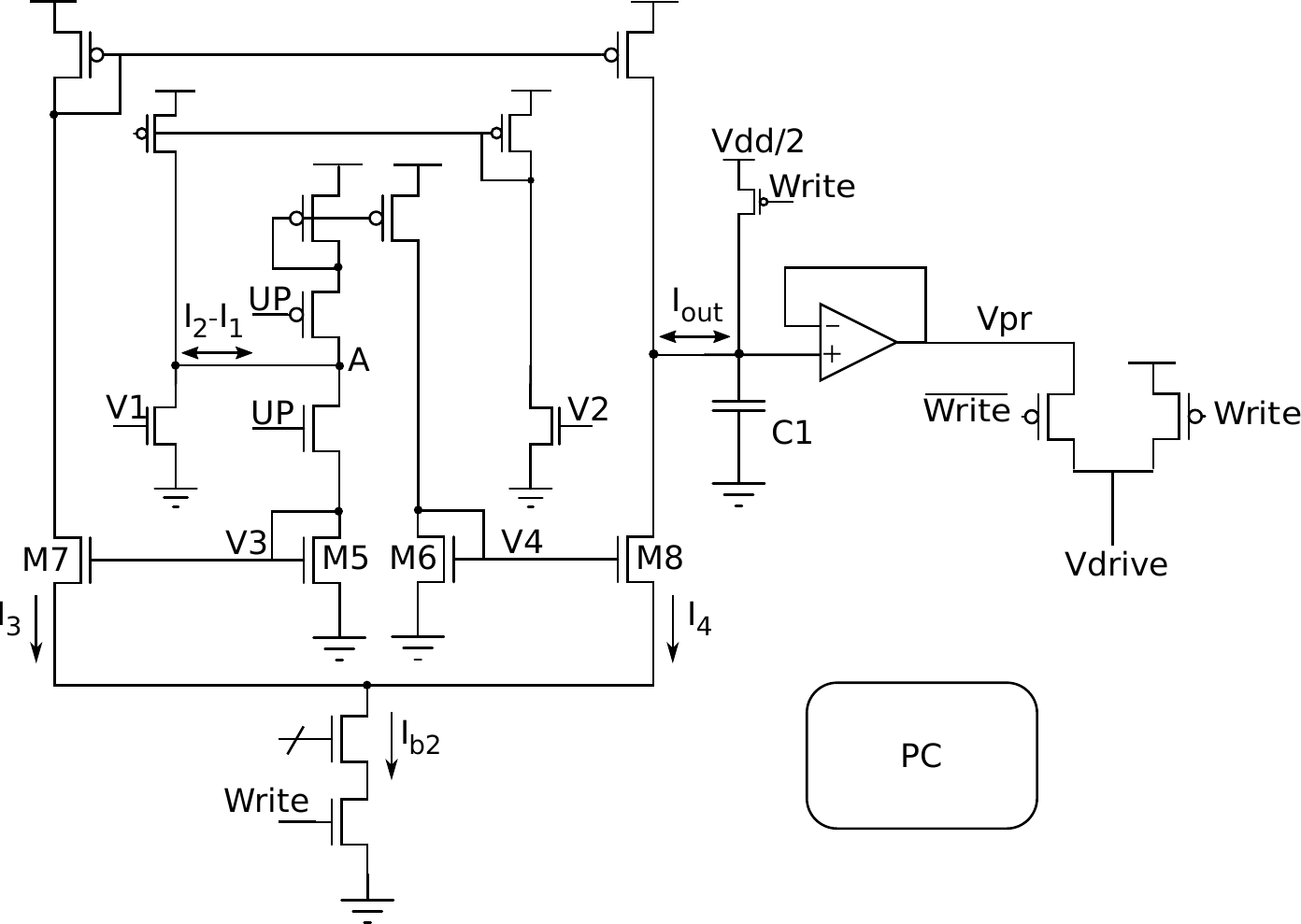}
 \caption{The Programming Circuit (PC) block used to generate the ramp used to program the memristors as a function of the error. The voltage signals \textsf{V1} and \textsf{V2} are obtained from Fig.~\ref{fig:bump}. Depending on the sign of the \textsf{UP} signal a rising or falling ramp is generated.}
 \label{fig:ramp}
\end{figure}

The circuit that produces this voltage ramp is shown in Fig.~\ref{fig:ramp}.
It is a global circuit shared by all the Memristive Synapse (MS) blocks of a neuron row (see PC block Fig.~\ref{fig:arch}).
The generation of the voltage ramp is triggered every time an input spike-event produces a \textsf{Write} pulse from the PS block of Fig.~\ref{fig:pulse_shaper}.
During this period the circuit is operational and receives as input the analog signals \textsf{V1}, \textsf{V2}, and the digital one \textsf{UP}. Given the subthreshold mode of operation, the output voltage signals of this circuit \textsf{V3} and \textsf{V4} can be expressed as:
\begin{equation}
  \begin{split}
    V3=\frac{U_T}{\kappa} \log(\frac{\Delta I}{I_0}) \quad \textrm{if} \quad \Delta I>0 
    \\
    V4=\frac{U_T}{\kappa} \log(\frac{-\Delta I}{I_0}) \quad \textrm{if} \quad \Delta I<0
  \end{split}
  \label{eq:logDiff}
\end{equation}

where $\Delta I$ is defined as $iTarget-iNeuron$, $k$ and $I_0$ are the process-dependent subthreshold slope factor and reverse biased leakage current respectively, and $U_T$ is the thermal voltage.

Now, to generate the desired ramp voltage, we need to convert the $(\mathsf{V3}-\mathsf{V4})$ voltage difference to a current that can charges/discharge a capacitor linearly. This is achieved by using a transconductance amplifier to produce the current  $I_{out}$:
\begin{equation}
  I_{out}=I_{b2} \tanh(\frac{\kappa}{2U_T}(V_3-V_4))
  \label{eq:Iout}
\end{equation}

It is safe to assume that the $tanh$ function of eq.~\eqref{eq:Iout} is operating in its linear region, since $V3$ and $V4$ are generated from $V1$ and $V2$ in circuits of Fig ~\ref{fig:bump} which operate in the subthreshold region. 
The ramp voltage $Vpr$ thus becomes:
\begin{equation}
  Vpr=\frac {V_{dd}}{2}\pm\frac{I_{out}}{C_1} \Delta t_{Write} \quad = \quad \frac {V_{dd}}{2}\pm\frac{I_{b2}}{2C_1} \log(\frac{\pm\Delta I}{I_0})\Delta t_{Write}
  \label{eqn:vpr}
\end{equation}
where $\Delta t_{Write} $ is the duration of the write-phase during which the memristors is programmed. The voltage $\frac {V_{dd}}{2}$ is the value to which the capacitor is pre-charged before and after the write-phase.

This voltage is applied to the memristive synapse that was stimulated by the input spike-event, using the polarity defined by the \textsf{UP} and \textsf{DN} signals produced by the Learning Block of the corresponding row.
As the ramp generator circuit is shared among all the synapses of a row, any other incoming spike-event received during the write-phase will be ignored. It has been shown that this assumption holds as long as the average rate of input spikes is slower than the write-phase ramp duration~\cite{Payvand_etal18}.

As the online learning proceeds and the neuron's mean activity approaches the target value, the magnitude of the current $I_{out}$ of the PC circuit (see Fig.~\ref{fig:ramp}) decreases and as a consequence the slope of the ramp decreases.
Since the probability of switching for the memristive devices is practically zero for voltages much lower than the ``nominal threshold voltage''~\cite{Ambrogio_etal16a}, this implementation induces a ``stop-learning'' zone in which no change is applied to the state of the devices.
It has been shown how this strategy of having a region of operation by which the weight-updates are disabled, when the learning error value decreases below a set threshold improves the stability of the learning process and the convergence properties of the network~\cite{Brader_etal07,Sheik_etal17,Baldassi_etal16}. Furthermore, this strategy has the important feature of enabling continuous time ``always on'' learning operations, without having to artificially separate the training phase from the test phase.

\begin{figure}
  \begin{subfigure}{0.45\textwidth}
    \centering \includegraphics[width=\textwidth]{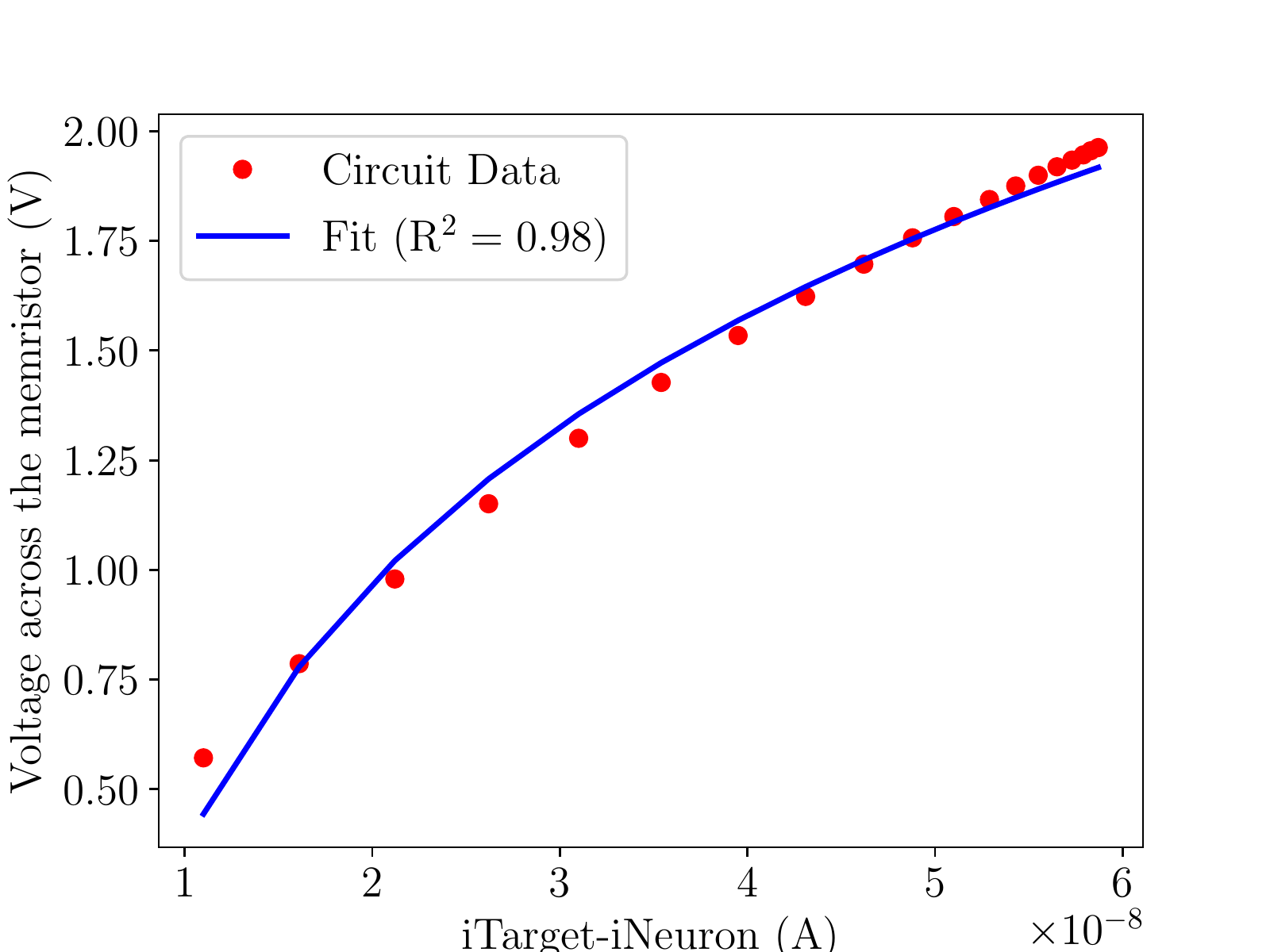}
    \subcaption{}
    \label{fig:ipos}
  \end{subfigure}
  \begin{subfigure}{0.45\textwidth}
    \centering
    \includegraphics[width=\textwidth]{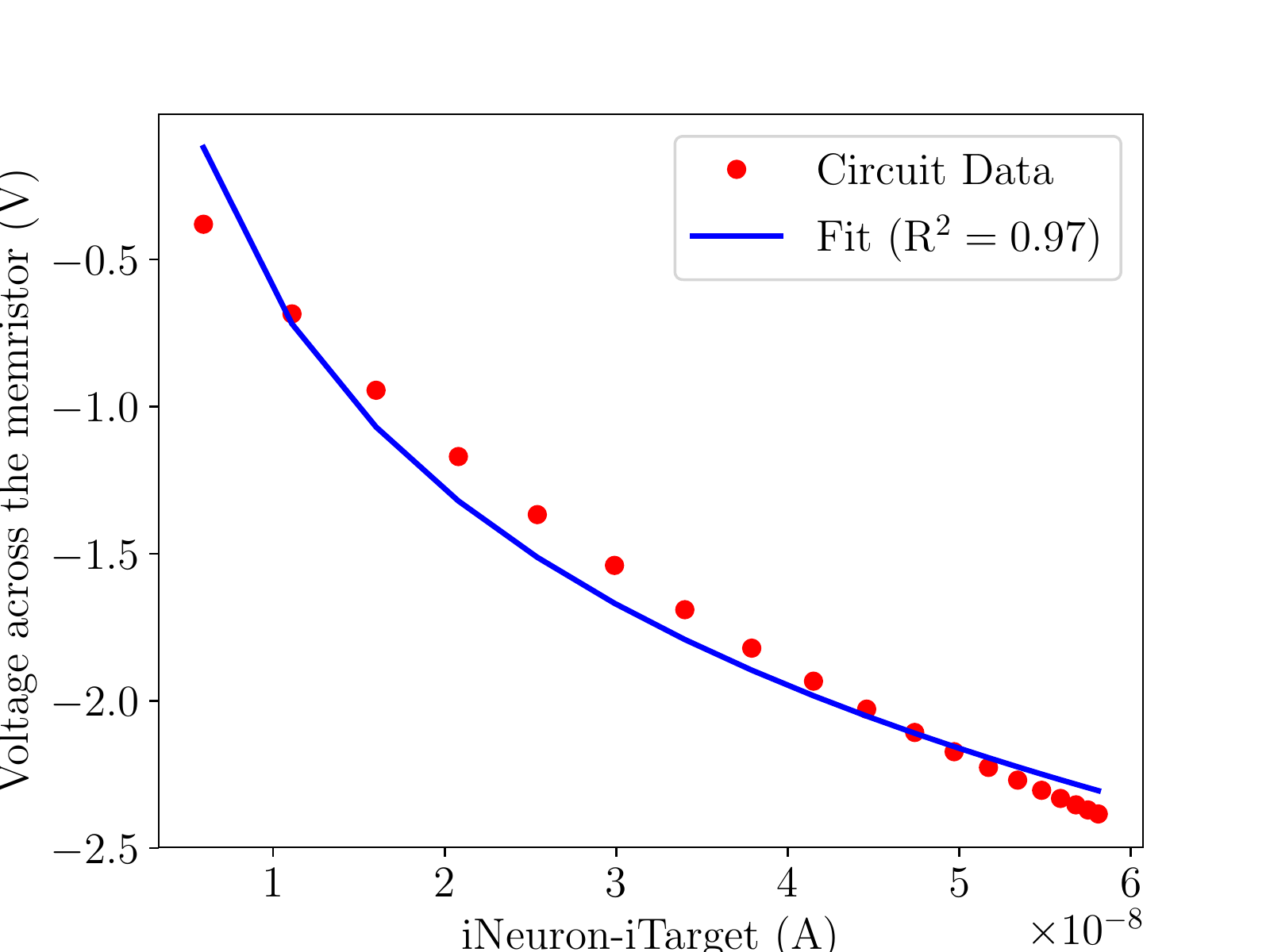}
    \subcaption{}
    \label{fig:ineg}
  \end{subfigure}
  \caption{Circuit simulation results. Voltage across the memristor is shown as a function of \textsf{iTarget-iNeuron} when the control signal \textsf{UP} is high (\subref{fig:ipos}), and  \textsf{iTarget-iNeuron} when the control signal \textsf{UP} is low (\subref{fig:ineg}). The circuit data is fitted with eq.~\eqref{eqn:vpr}. The term $R^2$ indicates the coefficient of determination, which is a statistical measure of how close the data are to the fitted line.}
  \label{fig:circuit_results}
\end{figure}

To validate the analysis presented above we carried out circuit simulations of both the Learning Block and the Programming Circuit of Fig.~\ref{fig:bump} and Fig.~\ref{fig:ramp} for a standard $0.18\,\mu$m CMOS process.
Figure~\ref{fig:circuit_results} shows the circuit simulation results, for both cases of the error signal $\Delta I$ greater and less than zero. The plots show also the fit of eq.~\eqref{eqn:vpr} with the data for $I_{b1}=50nA$, $I_{b2}=100nA$, $C_1=300fF$ and $\Delta t=10 \mu secs$. As depicted in the figures, the circuit outputs closely match the fits.

\section{System-level behavioral simulations}
\label{sec:sims}

To evaluate the effects of various sources of variability on the performance of the network and circuits proposed we carried out system-level behavioral simulations of the network, applied to a linear classification task using the MNIST hand-written digit data-set, comprising a training set for the learning phase and a test set for the validation phase. We compared the network performance on the test set after training on the training set in four cases:
\begin{enumerate}
\item Rate-based neural network with floating point synaptic precision trained by standard gradient-descent method as a baseline for comparing the accuracy of the network.
\item Spiking neural network with ideal binary devices trained by probabilistic gradient descent (as explained in Section~\ref{sec:stoch}).
\item Spiking neural network with non-ideal binary devices having high variability in their resistance value (20 \% of standard deviation) trained by probabilistic gradient descent.
\item Spiking neural network with non-ideal binary devices of item 3, whose variations are suppressed using the  variability reduction circuit presented in Section~\ref{sec:mism-norm}, and trained by probabilistic gradient descent.
\end{enumerate}

To compare the network to previously published results, we used a configuration analogous to the setup presented in the work of Bill and Legenstein~\cite{Bill_Legenstein14}, who used a model of memristive elements in an unsupervised Winner-take-all network to learn digit prototypes for digits zero to four. A downscaled network of this kind has been partially verified in hardware recently~\cite{Serb_etal16}. This setup is also comparable to other setups for previous simulations done by our group~\cite{Nair_etal17,Payvand_etal18}.

We carried out spiking neural network simulations using the Brian2 simulator~\cite{Goodman_Brette09} and neuron model equations that match the transfer function of the silicon neuron circuits~\cite{Chicca_etal14} and DPI filters~\cite{Bartolozzi_Indiveri07a} used in the architecture. In these simulations, we combine for the first time a stochastic learning algorithm with a variability compensation method. Both are based on different variability characteristics of memristors: The stochastic learning algorithm uses the cycle-to-cycle variability in the switching probability of a memristor for a given voltage ramp, the variability compensation addresses the device-to-device (and cycle-to-cycle) variability in conductance level of a memristor.  

The gray-level MNIST input images were re-scaled to image sizes of $24 \times 24$ and their pixel values were converted to Poisson spike trains with a mean firing rate proportional to the pixel intensity. To obtain higher resolution effective connections from each input pixel while using binary synaptic elements we encoded the pixel values with multiple instances of spiking neurons. Specifically, each pixel was associated to a number $n_c$ of spiking neurons in the input layer, that stimulated a corresponding number of synaptic elements of a target ``compound synapse'' (comprising $n_c$ devices instead of two) in the network output recognition layer. In this way, the synaptic connection strengths have $2\cdot n_c$ effective levels, instead of two.
The total number of neurons in the input layer is therefore $n_c (24 \times 24)$. The output recognition layer is composed of five read-out neurons (one for each digit type zero to four), each of which comprises a row of $(24 \times 24)$ compound synapses, with each compound synapse containing $n_c$ memristive devices.

The neuromorphic architecture used in these system level behavioral simulations is the one described in Section~\ref{sec:arch}. The parameters used to encode the synaptic weights are either two precise discrete values (with no variability), in the case of idealized synaptic elements, or are random numbers that follow a bi-modal distribution based on measured data from memristive device properties, as given in Fig.~\ref{fig:mismatch_res_var}~and~\ref{fig:mismatch_res_var10}. To implement the learning strategy described in Section~\ref{sec:stoch} we model the effect of the ramp generator on the synaptic conductance as a stochastic binary update, using the switching probabilities defined in Section~\ref{sec:stoch}.
The learning block of each output neuron receives inputs from two sources: from the filter that measures the average firing rate of the neuron itself \textsf{iNeuron}, and from external teacher neurons that provide a desired average current \textsf{iTarget}. In the protocol used, large \textsf{iTarget} values indicate that the neuron should learn to be active for the given input pattern (see also Fig.~\ref{fig:bump}), while low \textsf{iTarget} values indicate that the neuron should learn to ignore the input pattern.

The network is initialized by sampling synaptic weights from appropriate distributions given in Fig.~\ref{fig:mismatch_res_var} and Fig.~\ref{fig:mismatch_res_var10}.
Note that we assume that the memristive devices have already been formed and are ready for read and write operations.
Training the network is achieved by presenting 10000 randomly chosen digits from the training set along with the appropriate teacher signals. Each image is presented for 100 ms while the learning circuits tune the synaptic weights. After this, the performance of the network is evaluated on 5000 further digits (randomly drawn from the test set).

\begin{figure}
 \centering
 \includegraphics[width=0.45\textwidth]{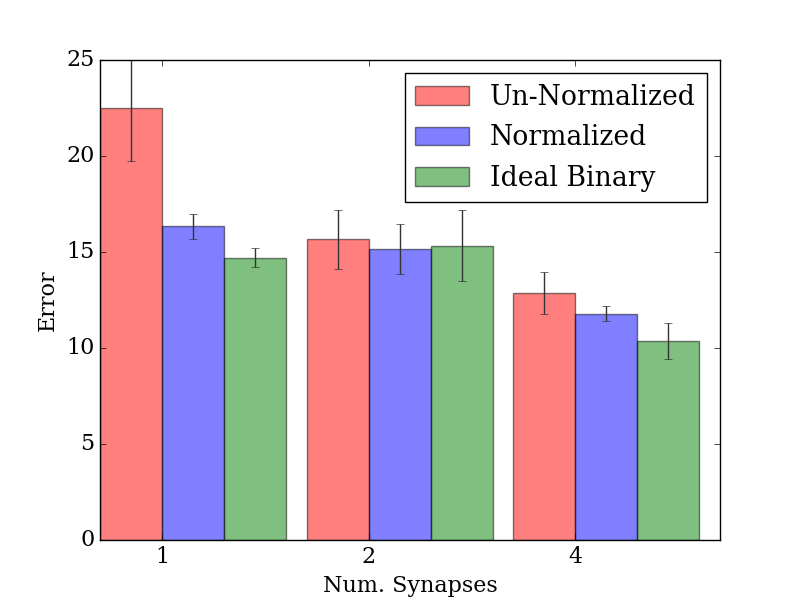}
 \caption{Test set error on MNIST digits 0-4 vs. number of synapses per input pixel ($n_c$). On/Off ratio in the memristor model is 2 (see Fig. \ref{fig:tf}).}
 \label{fig:mnist1}
\end{figure}

\begin{figure}
 \centering
 \includegraphics[width=0.45\textwidth]{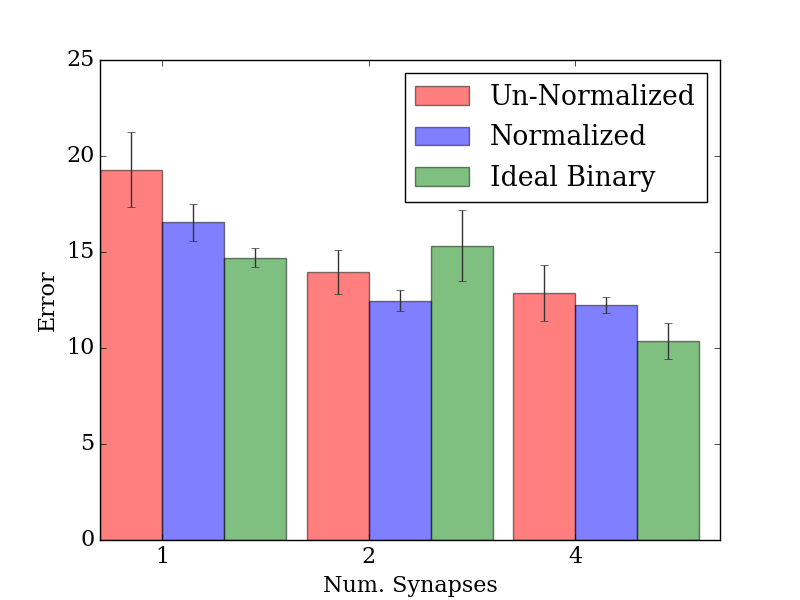}
 \caption{Test set error on MNIST digits 0-4 vs. number of synapses per input pixel ($n_c$). On/Off ratio in the memristor model is 10 (see Fig. \ref{fig:tf10}).}
 \label{fig:mnist2}
\end{figure}

To evaluate the performance of the network, namely the classification accuracy, we chose as network output the index of the output neuron that spiked with the highest firing rate during the input presentation and compared it's identity to the label of the pattern provided in input. If more than one output neuron spiked, the neuron that spiked the most was chosen as the one encoding the learned label.

Figures~\ref{fig:mnist1} and~\ref{fig:mnist2} show the performance of the proposed architecture. As a base-line comparison (that we expect to upper bound the performance of this setup) we also trained a standard linear classifier with 32-bit floating point synaptic elements and 32-bit rate based neurons using stochastic gradient descent~\cite{Bishop06}. This baseline reaches circa $2.9 \% \pm .1\%$ test set error.
The discrepancy to the circa $10 \%$ error of our best simulation, can be explained by the low resolution of synaptic memory, the single bit communication channels of spiking neurons and the lossy input encoding in Poisson spike trains. An intermediary idealized setup, only controlling for memristive conductance variability, but incorporating other non-idealities is given by the `ideal binary' simulations (see the green bars in Fig.~\ref{fig:mnist1} and~\ref{fig:mnist2}).
The network simulations with different types of synapse models (i.e., basic un-normalized linear conversion case, and current-normalizer conversion case) show how the normalization circuit  decreases the classification error overall. By comparing the error-bars on the un-normalized (red bars) and normalized (blue bars) simulation results in Fig.~\ref{fig:mnist1} and~\ref{fig:mnist2}, it is evident how the normalization circuit decreases also the variance in the error. We speculate that the reason for this is the more stable update size of the normalized setup.

\begin{figure*}[!htbp]
 \centering
 \includegraphics[width=\textwidth]{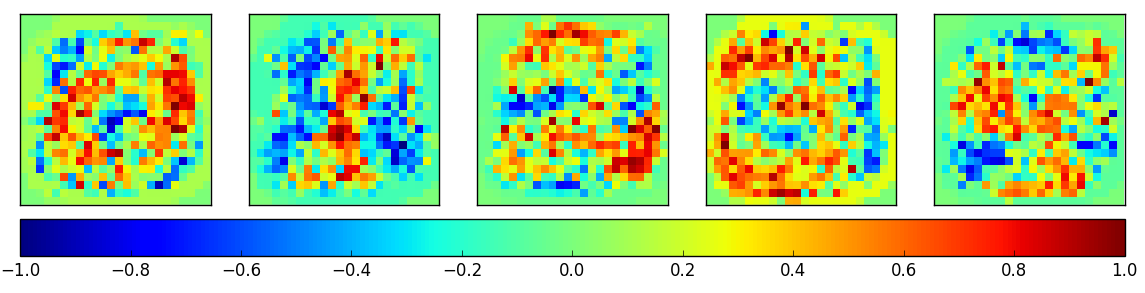}
 \caption{Example receptive fields (arbitrary units) learned by the network with $n_c=4$. Note that these are discriminatory features, not digit prototypes.}
 \label{fig:recFields}
\end{figure*}

Figure~\ref{fig:recFields} shows examples of synaptic weight matrices of the five different neurons that were trained to recognize the five different digits, for the case in which $n_c=4$. These synaptic weight matrices can be interpreted as ``receptive fields'' of the trained neurons which correspond to the best discriminatory features (e.g., positive weights for prototypes of the digit the neurons are supposed to classify intermixed with negative weights for the digits that they are supposed to ignore).

Overall these simulations show that changes on the behavioral level of a small neural network can be influenced by low-level characteristics of the building blocks of the neurons that comprise it. Specifically we have shown that the probabilistic switching behavior of memristors can be used as a powerful computational primitive in a learning setting, and that variability in conductance levels of memristors can be effectively (in the sense of high-level performance) mitigated by appropriate normalization with a compact circuit.

\section{Discussion}
\label{sec:discussion}
\subsection{Supporting different modes of memristive device operation}
\label{sec:supp-diff-modes}
Although in this paper we focus on the use of memristive devices as binary elements, the architecture proposed can potentially support the full spectrum of memristive behaviors that has been reported in the literature: 
\begin{enumerate}
\item Stochastic binary~\cite{Gaba_etal13}
\item Multiple binary devices in parallel (compound synapse)~\cite{Boybat_etal18,Serb_etal16,Gaba_etal13}
\item Stochastic multiple discrete levels~\cite{Serb_etal16,Stathopoulos_etal17} 
\item Almost analog~\cite{frascaroli_etal18,chang_etal11,prezioso_etal15b}
\end{enumerate}

In the case of binary synapses, we showed how the proposed stochastic learning circuits enable the architecture to achieve acceptable performance on the MNIST test bench. The system-level behavioral simulations demonstrated that the use of compound synapses improves the classification performance, and quantified the improvement factors.

It has been shown in the literature~\cite{frascaroli_etal18,chang_etal11}, how gradual conductance modulation of memristive devices can be observed when pulses are applied for a short amount of time. Under these conditions controlling the number of pulses applied to the device can be used as a way to tune the desired conductance values. The architecture proposed can support this regime of operation by appropriately setting the the pulse height and/or duration via the \textsf{LB} and \textsf{PC} blocks of Section~\ref{sec:stoch}. The same circuits can be extended to produce a tunable number of short pulse sequences by enabling a ring oscillator for the desired duration. This latter strategy would allow us to implement learning with gradual changes, rather than binary probabilistic one, by encoding the desired change in weight $\Delta w$ with the number of pulses generated by the ring oscillator.
It is worth noting that the same memristive device can be tuned to behave as a binary one or multi-level one by adopting different biasing and operating conditions~\cite{frascaroli_etal18}. For example, even for a fixed set-voltage, it is possible to operate the same device in the binary or analog region by changing the length of the \textsf{Write} pulse in the \textsf{PS} block of Fig.~\ref{fig:pulse_shaper}: longer pulses will drive the device into the binary mode, while shorter ones will exhibit more of an analog behavior.

\subsection{Exploiting device mismatch and variability to improve classification accuracy}
\label{sec:impr-accur-with}

In this paper we have presented analog CMOS circuits that can be interfaced to memristive devices to \emph{mitigate} the effect of their device variability. A remarkable feature of the use of analog CMOS circuits used to implement also synapse and neuron dynamics is the fact that their device mismatch non-idealities can be \emph{exploited} to improve the network classification performance. Indeed, device mismatch across multiple memristive synapses and silicon neurons, the very phenomenon that decreases the classification performance of one single binary classifier (e.g., one Perceptron or neuron row of Fig.~\ref{fig:arch}) and that engineers tend to minimize with brute-force approaches, can be embraced to build highly accurate classifiers composed of \emph{ensembles} of single ones. This can be demonstrated by the theory of \emph{ensemble learning}. There are two broad classes of algorithms that fall in the category of ensemble learning: Bagging and Boosting.
\begin{description}
\item[Bagging] or bootstrap aggregating is an averaging technique proposed by Breiman~\cite{Breiman_96bagging} where a collection of M classifiers are trained on M equally-sized subsets of the full training set created with replacement. The predictions made by the ensemble of M classifiers are then averaged to make the final prediction. 
\item[Boosting] is a technique that uses a collection of un-correlated weak classifiers (whose accuracy is only slightly better than chance) to build a strong-classifier (whose prediction error can be made arbitrarily small)~\cite{Schapire_90boosting}. One of the most popular variants of the approach is called the AdaBoost algorithm~\cite{Freund_etal97adaboost}. Unlike the bagging approach, every weak classifier in the ensemble is exposed to the full training data, where each sample is associated with an observation weight during training. For training the first classifier, the weights are kept equal for every training sample. When training the second classifier, the sample weights are adjusted such that the misclassified samples by the first classifier have a higher weight. A weight is also assigned to each classifier based on its prediction accuracy. This process is continued till the desired number of weak classifiers are generated. The final prediction from the ensemble is a weighted sum of the weak-classifier predictions.
\end{description}

These ensemble learning principles can indeed be applied to the neuromorphic architecture proposed in Section~\ref{sec:arch} to asymptotically improve the accuracy of the system. In particular, the Bagging approach is immediately applicable to the system, by simply sending the same input patterns to multiple neuron rows and training ensembles of neurons to recognize the same class. The variability in the synapse and neuron circuits is already sufficient to make sure that each neuron acting as a ``weak'' binary classifier behaves in a way that is different from the other ones belonging to the same ensemble.
However, to truly ensure that the weak classifiers are fully independent it would be sufficient to train each neuron of the same ensemble with input patterns that represent different sub data-sets of the original training data set. This has indeed already been demonstrated with pure CMOS based architectures of the type proposed in this paper, by using different random connectivity patterns for each weak classifier of the ensemble~\cite{Qiao_etal15,Corradi_etal14}.

The boosting approach, promises to yield even better results. However the constraints on choosing which weights to change might lead to the adoption of extra control modules per neuron that require too large or complex overhead circuits and could result to be prohibitive for realistic compact chip designs.

\subsection{Cross-bars versus addressable arrays}
\label{sec:xbars}

The nano-scale footprint of memristors~\cite{Yang_etal13,pi_etal13cross,govoreanu_etal13vacancy} is an important feature which can enable ultra dense memory capacity~\cite{Payvand_etal15,Stathopoulos_etal17}. To exploit this extremely low footprint to its full extent, dense cross-bar arrays have been reportedly implemented and proposed as in-memory computing neural network engines~\cite{Kim_etal12,Prezioso_etal15,Ambrogio_etal18}.
However, although the development of dense cross-bars is extremely important for the scaling of technology, there are many challenges associated with their use in neuromorphic architectures both from fabrication and circuits point of view.
For example, it is not clear how much passive cross-bar arrays can be scaled up to larger sizes, due to sneak path and cross-talk issues~\cite{Kim_etal12}.
Even in the case of cross-bar arrays with active elements such as 1T-1R (one-transistor and one-memristor) or memristive devices with embedded ``selectors'' used to avoid the sneak-path problem, issues such as the line resistance, reproducibility, and overhead size of external encoder and decoder CMOS circuits~\cite{Sandrini_etal16} are yet to be satisfactorily addressed.
Alternatively, one can decide to forgo the cross-bar approach of very high density arrangements of basic 1R or 1T-1R elements, and design addressable arrays of more complex synapses that comprise multiple transistors and multiple memristive devices per synapse, to try and capitalize on the many other useful features of memristive devices (in addition to their compact size), such as non-volatility, state-dependence, complex physics that can be exploited to emulate the complex molecular properties of biological synapses, complex dynamics, and stochastic switching behavior.
The architecture we propose represents an intermediate approach that comprises two memristive devices per synapse and two select switches.
This design was proposed to allow maximum flexibility in exploring the properties of different types of memristive memory devices, but it could be made even more dense by replacing the transistors currently used to switch between read-mode and write-mode with embedded selectors and modulating the amplitude of the \textsf{Vdrive} line of Fig.~\ref{fig:synapse} to operate the device only in read-mode or in both read- and write-mode, thanks to the fact that the voltage set at the terminals of the memristive devices is a ramp that can cover both ranges of operation.
However, while large-scale in-memory computing cross-bar arrays of this type may solve the memory-bottleneck problem~\cite{Backus78,Indiveri_Liu15}, they would still be crippled by an Input/Output (I/O) bottleneck problem due to the constraint that while one synapse is being operated in its write-mode (which could last micro-seconds), no other synapse of the same row could be stimulated.
By incorporating the \textsf{PS} and \textsf{NC} blocks of Fig.~\ref{fig:arch} in the \textsf{MS} blocks, this addressable array architecture would definitely lose the benefit of high-density synapses, but would dramatically increase the bandwidth of its input Address-Events (e.g., with each I/O operation lasting nano-seconds), as each synapse element would become independent from the others and multiple synapses would be able to safely operate in read- or write-mode in parallel.
Once the choice is made to forgo the density benefit, adding further transistors for example to implement local non-linear dynamics, such as short-term plasticity~\cite{Boegerhausen_etal03}, or homeostatic synaptic scaling mechanisms~\cite{Qiao_etal17}, or more complex learning mechanisms~\cite{Huayaney_etal16} to improve the performance of the overall neuromorphic computing system would become easily realizable. 

\section{Conclusions}
\label{sec:conclusions}

We presented an effort to design and combine a suite of computational techniques for constructing a trainable neuromorphic platform that supports the use of a wide variety of memristive devices. We showed that variability of the memristive devices and mismatch in CMOS circuits can be on one hand reduced by circuit techniques, and can on the other hand be exploited as a feature for training and computation.  We described the architecture of a neuromorphic platform that can implement stochastic training exploiting the switching properties of memristive devices and validated the approach with system-level behavioral simulations for a linear classification task, using the MNIST data-set. 

The proposed neuromorphic computing architecture supports continuous-time always-on on-chip learning, and continuously streams output spikes to the AER output block. By routing output address-events via either off-chip or on-chip asynchronous AER routing schemes and circuits~\cite{Moradi_etal15,Fasnacht_Indiveri11,Serrano-Gotarredona_etal09},  these architectures support scaling by tiling them either across multiple chips, or on multiple cores within a multi-core device. Examples of multi-core neuromorphic computing systems based on the AER protocol have been recently proposed~\cite{Davies_etal18,Moradi_etal17,Park_etal16,Merolla_etal14a}, however none have been implemented so far using memristive devices, and exploiting their intrinsic properties to implement probabilistic learning. 

\section*{Acknowledgements}
This work is supported by SNSF grant number CRSII2\_160756. We acknowledge also funding from the ``Internationalization Fund of the FZ-Juelich'' for the project ``NeuroCode''.





\bibliography{faraday18}

\providecommand*{\mcitethebibliography}{\thebibliography}
\csname @ifundefined\endcsname{endmcitethebibliography}
{\let\endmcitethebibliography\endthebibliography}{}
\begin{mcitethebibliography}{82}
\providecommand*{\natexlab}[1]{#1}
\providecommand*{\mciteSetBstSublistMode}[1]{}
\providecommand*{\mciteSetBstMaxWidthForm}[2]{}
\providecommand*{\mciteBstWouldAddEndPuncttrue}
  {\def\EndOfBibitem{\unskip.}}
\providecommand*{\mciteBstWouldAddEndPunctfalse}
  {\let\EndOfBibitem\relax}
\providecommand*{\mciteSetBstMidEndSepPunct}[3]{}
\providecommand*{\mciteSetBstSublistLabelBeginEnd}[3]{}
\providecommand*{\EndOfBibitem}{}
\mciteSetBstSublistMode{f}
\mciteSetBstMaxWidthForm{subitem}
{(\emph{\alph{mcitesubitemcount}})}
\mciteSetBstSublistLabelBeginEnd{\mcitemaxwidthsubitemform\space}
{\relax}{\relax}

\bibitem[Chicca \emph{et~al.}(2014)Chicca, Stefanini, Bartolozzi, and
  Indiveri]{Chicca_etal14}
E.~Chicca, F.~Stefanini, C.~Bartolozzi and G.~Indiveri, \emph{Proceedings of
  the {IEEE}}, 2014, \textbf{102}, 1367--1388\relax
\mciteBstWouldAddEndPuncttrue
\mciteSetBstMidEndSepPunct{\mcitedefaultmidpunct}
{\mcitedefaultendpunct}{\mcitedefaultseppunct}\relax
\EndOfBibitem
\bibitem[Park \emph{et~al.}(2014)Park, Ha, Yu, Neftci, and
  Cauwenberghs]{Park_etal14}
J.~Park, S.~Ha, T.~Yu, E.~Neftci and G.~Cauwenberghs, Biomedical Circuits and
  Systems Conference (BioCAS), 2014 IEEE, 2014, pp. 675--678\relax
\mciteBstWouldAddEndPuncttrue
\mciteSetBstMidEndSepPunct{\mcitedefaultmidpunct}
{\mcitedefaultendpunct}{\mcitedefaultseppunct}\relax
\EndOfBibitem
\bibitem[Furber \emph{et~al.}(2014)Furber, Galluppi, Temple, and
  Plana]{Furber_etal14}
S.~Furber, F.~Galluppi, S.~Temple and L.~Plana, \emph{Proceedings of the IEEE},
  2014, \textbf{102}, 652--665\relax
\mciteBstWouldAddEndPuncttrue
\mciteSetBstMidEndSepPunct{\mcitedefaultmidpunct}
{\mcitedefaultendpunct}{\mcitedefaultseppunct}\relax
\EndOfBibitem
\bibitem[Benjamin \emph{et~al.}(2014)Benjamin, Gao, McQuinn, Choudhary,
  Chandrasekaran, Bussat, Alvarez-Icaza, Arthur, Merolla, and
  Boahen]{Benjamin_etal14}
B.~V. Benjamin, P.~Gao, E.~McQuinn, S.~Choudhary, A.~R. Chandrasekaran,
  J.~Bussat, R.~Alvarez-Icaza, J.~Arthur, P.~Merolla and K.~Boahen,
  \emph{Proceedings of the {IEEE}}, 2014, \textbf{102}, 699--716\relax
\mciteBstWouldAddEndPuncttrue
\mciteSetBstMidEndSepPunct{\mcitedefaultmidpunct}
{\mcitedefaultendpunct}{\mcitedefaultseppunct}\relax
\EndOfBibitem
\bibitem[Merolla \emph{et~al.}(2014)Merolla, Arthur, Alvarez, Bussat, and
  Boahen]{Merolla_etal14}
P.~Merolla, J.~Arthur, R.~Alvarez, J.-M. Bussat and K.~Boahen, \emph{Circuits
  and Systems I: Regular Papers, {IEEE} Transactions on}, 2014, \textbf{61},
  820--833\relax
\mciteBstWouldAddEndPuncttrue
\mciteSetBstMidEndSepPunct{\mcitedefaultmidpunct}
{\mcitedefaultendpunct}{\mcitedefaultseppunct}\relax
\EndOfBibitem
\bibitem[Mitra \emph{et~al.}(2009)Mitra, Fusi, and Indiveri]{Mitra_etal09}
S.~Mitra, S.~Fusi and G.~Indiveri, \emph{Biomedical Circuits and Systems,
  {IEEE} Transactions on}, 2009, \textbf{3}, 32--42\relax
\mciteBstWouldAddEndPuncttrue
\mciteSetBstMidEndSepPunct{\mcitedefaultmidpunct}
{\mcitedefaultendpunct}{\mcitedefaultseppunct}\relax
\EndOfBibitem
\bibitem[Qiao \emph{et~al.}(2015)Qiao, Mostafa, Corradi, Osswald, Stefanini,
  Sumislawska, and Indiveri]{Qiao_etal15}
N.~Qiao, H.~Mostafa, F.~Corradi, M.~Osswald, F.~Stefanini, D.~Sumislawska and
  G.~Indiveri, \emph{Frontiers in Neuroscience}, 2015, \textbf{9}, 1--17\relax
\mciteBstWouldAddEndPuncttrue
\mciteSetBstMidEndSepPunct{\mcitedefaultmidpunct}
{\mcitedefaultendpunct}{\mcitedefaultseppunct}\relax
\EndOfBibitem
\bibitem[Moradi \emph{et~al.}(2017)Moradi, Qiao, Stefanini, and
  Indiveri]{Moradi_etal17}
S.~Moradi, N.~Qiao, F.~Stefanini and G.~Indiveri, \emph{Biomedical Circuits and
  Systems, {IEEE} Transactions on}, 2017,  1--17\relax
\mciteBstWouldAddEndPuncttrue
\mciteSetBstMidEndSepPunct{\mcitedefaultmidpunct}
{\mcitedefaultendpunct}{\mcitedefaultseppunct}\relax
\EndOfBibitem
\bibitem[Davies \emph{et~al.}(2018)Davies, Srinivasa, Lin, Chinya, Cao, Choday,
  Dimou, Joshi, Imam, Jain, Liao, Lin, Lines, Liu, Mathaikutty, McCoy, Paul,
  Tse, Venkataramanan, Weng, Wild, Yang, and Wang]{Davies_etal18}
M.~Davies, N.~Srinivasa, T.~H. Lin, G.~Chinya, Y.~Cao, S.~H. Choday, G.~Dimou,
  P.~Joshi, N.~Imam, S.~Jain, Y.~Liao, C.~K. Lin, A.~Lines, R.~Liu,
  D.~Mathaikutty, S.~McCoy, A.~Paul, J.~Tse, G.~Venkataramanan, Y.~H. Weng,
  A.~Wild, Y.~Yang and H.~Wang, \emph{{IEEE} Micro}, 2018, \textbf{38},
  82--99\relax
\mciteBstWouldAddEndPuncttrue
\mciteSetBstMidEndSepPunct{\mcitedefaultmidpunct}
{\mcitedefaultendpunct}{\mcitedefaultseppunct}\relax
\EndOfBibitem
\bibitem[Backus(1978)]{Backus78}
J.~Backus, \emph{Communications of the ACM}, 1978, \textbf{21}, 613--641\relax
\mciteBstWouldAddEndPuncttrue
\mciteSetBstMidEndSepPunct{\mcitedefaultmidpunct}
{\mcitedefaultendpunct}{\mcitedefaultseppunct}\relax
\EndOfBibitem
\bibitem[Indiveri and Liu(2015)]{Indiveri_Liu15}
G.~Indiveri and S.-C. Liu, \emph{Proceedings of the {IEEE}}, 2015,
  \textbf{103}, 1379--1397\relax
\mciteBstWouldAddEndPuncttrue
\mciteSetBstMidEndSepPunct{\mcitedefaultmidpunct}
{\mcitedefaultendpunct}{\mcitedefaultseppunct}\relax
\EndOfBibitem
\bibitem[Boybat \emph{et~al.}(2018)Boybat, Gallo, Moraitis, Parnell, Tuma,
  Rajendran, Leblebici, Sebastian, Eleftheriou,\emph{et~al.}]{Boybat_etal18}
I.~Boybat, M.~L. Gallo, T.~Moraitis, T.~Parnell, T.~Tuma, B.~Rajendran,
  Y.~Leblebici, A.~Sebastian, E.~Eleftheriou \emph{et~al.}, \emph{Nature
  communications}, 2018, \textbf{9}, 2514\relax
\mciteBstWouldAddEndPuncttrue
\mciteSetBstMidEndSepPunct{\mcitedefaultmidpunct}
{\mcitedefaultendpunct}{\mcitedefaultseppunct}\relax
\EndOfBibitem
\bibitem[Li \emph{et~al.}(2018)Li, Belkin, Li, Yan, Hu, Ge, Jiang, Montgomery,
  Lin, Wang, Song, Strachan, Barnell, Wu, Williams, Yang, and Xia]{Li_etal18}
C.~Li, D.~Belkin, Y.~Li, P.~Yan, M.~Hu, N.~Ge, H.~Jiang, E.~Montgomery, P.~Lin,
  Z.~Wang, W.~Song, J.~P. Strachan, M.~Barnell, Q.~Wu, R.~S. Williams, J.~J.
  Yang and Q.~Xia, \emph{Nature Communications}, 2018, \textbf{9}, 1--8\relax
\mciteBstWouldAddEndPuncttrue
\mciteSetBstMidEndSepPunct{\mcitedefaultmidpunct}
{\mcitedefaultendpunct}{\mcitedefaultseppunct}\relax
\EndOfBibitem
\bibitem[Ambrogio \emph{et~al.}(2018)Ambrogio, Narayanan, Tsai, Shelby, Boybat,
  di~Nolfo, Sidler, Giordano, Bodini, Farinha, Killeen, Cheng, Jaoudi, and
  Burr]{Ambrogio_etal18}
S.~Ambrogio, P.~Narayanan, H.~Tsai, R.~M. Shelby, I.~Boybat, C.~di~Nolfo,
  S.~Sidler, M.~Giordano, M.~Bodini, N.~C.~P. Farinha, B.~Killeen, C.~Cheng,
  Y.~Jaoudi and G.~W. Burr, \emph{Nature}, 2018, \textbf{558}, 60--67\relax
\mciteBstWouldAddEndPuncttrue
\mciteSetBstMidEndSepPunct{\mcitedefaultmidpunct}
{\mcitedefaultendpunct}{\mcitedefaultseppunct}\relax
\EndOfBibitem
\bibitem[Likharev \emph{et~al.}(2003)Likharev, Mayr, Muckra, and
  T{\"u}rel]{Likharev_etal03}
K.~Likharev, A.~Mayr, I.~Muckra and {\"O}.~T{\"u}rel, \emph{Annals of the New
  York Academy of Sciences}, 2003, \textbf{1006}, 146--163\relax
\mciteBstWouldAddEndPuncttrue
\mciteSetBstMidEndSepPunct{\mcitedefaultmidpunct}
{\mcitedefaultendpunct}{\mcitedefaultseppunct}\relax
\EndOfBibitem
\bibitem[Linn \emph{et~al.}(2010)Linn, Rosezin, K{\"u}geler, and
  Waser]{Linn_etal10}
E.~Linn, R.~Rosezin, C.~K{\"u}geler and R.~Waser, \emph{Nature materials},
  2010, \textbf{9}, 403--406\relax
\mciteBstWouldAddEndPuncttrue
\mciteSetBstMidEndSepPunct{\mcitedefaultmidpunct}
{\mcitedefaultendpunct}{\mcitedefaultseppunct}\relax
\EndOfBibitem
\bibitem[Kim \emph{et~al.}(2012)Kim, Gaba, Wheeler, Cruz-Albrecht, Hussain,
  Srinivasa, and Lu]{Kim_etal12}
K.~Kim, S.~Gaba, D.~Wheeler, J.~Cruz-Albrecht, T.~Hussain, N.~Srinivasa and
  W.~Lu, \emph{Nano letters}, 2012, \textbf{12}, 389--395\relax
\mciteBstWouldAddEndPuncttrue
\mciteSetBstMidEndSepPunct{\mcitedefaultmidpunct}
{\mcitedefaultendpunct}{\mcitedefaultseppunct}\relax
\EndOfBibitem
\bibitem[Prezioso \emph{et~al.}(2015)Prezioso, Merrikh-Bayat, Hoskins, Adam,
  Likharev, and Strukov]{Prezioso_etal15}
M.~Prezioso, F.~Merrikh-Bayat, B.~Hoskins, G.~Adam, K.~K. Likharev and D.~B.
  Strukov, \emph{Nature}, 2015, \textbf{521}, 61--64\relax
\mciteBstWouldAddEndPuncttrue
\mciteSetBstMidEndSepPunct{\mcitedefaultmidpunct}
{\mcitedefaultendpunct}{\mcitedefaultseppunct}\relax
\EndOfBibitem
\bibitem[Sandrini \emph{et~al.}(2016)Sandrini, Barlas, Thammasack, Demirci,
  De~Marchi, Sacchetto, Gaillardon, De~Micheli, and Leblebici]{Sandrini_etal16}
J.~Sandrini, M.~Barlas, M.~Thammasack, T.~Demirci, M.~De~Marchi, D.~Sacchetto,
  P.-E. Gaillardon, G.~De~Micheli and Y.~Leblebici, \emph{IEEE Journal on
  Emerging and Selected Topics in Circuits and Systems}, 2016, \textbf{6},
  339--351\relax
\mciteBstWouldAddEndPuncttrue
\mciteSetBstMidEndSepPunct{\mcitedefaultmidpunct}
{\mcitedefaultendpunct}{\mcitedefaultseppunct}\relax
\EndOfBibitem
\bibitem[Merolla \emph{et~al.}(2014)Merolla, Arthur, Alvarez-Icaza, Cassidy,
  Sawada, Akopyan, Jackson, Imam, Guo, Nakamura, Brezzo, Vo, Esser, Appuswamy,
  Taba, Amir, Flickner, Risk, Manohar, and Modha]{Merolla_etal14a}
P.~A. Merolla, J.~V. Arthur, R.~Alvarez-Icaza, A.~S. Cassidy, J.~Sawada,
  F.~Akopyan, B.~L. Jackson, N.~Imam, C.~Guo, Y.~Nakamura, B.~Brezzo, I.~Vo,
  S.~K. Esser, R.~Appuswamy, B.~Taba, A.~Amir, M.~D. Flickner, W.~P. Risk,
  R.~Manohar and D.~S. Modha, \emph{Science}, 2014, \textbf{345},
  668--673\relax
\mciteBstWouldAddEndPuncttrue
\mciteSetBstMidEndSepPunct{\mcitedefaultmidpunct}
{\mcitedefaultendpunct}{\mcitedefaultseppunct}\relax
\EndOfBibitem
\bibitem[Payvand \emph{et~al.}(2015)Payvand, Madhavan, Lastras-Monta{\~n}o,
  Ghofrani, Rofeh, Cheng, Strukov, and Theogarajan]{Payvand_etal15}
M.~Payvand, A.~Madhavan, M.~A. Lastras-Monta{\~n}o, A.~Ghofrani, J.~Rofeh,
  K.-T. Cheng, D.~Strukov and L.~Theogarajan, Circuits and Systems (ISCAS),
  2015 IEEE International Symposium on, 2015, pp. 1378--1381\relax
\mciteBstWouldAddEndPuncttrue
\mciteSetBstMidEndSepPunct{\mcitedefaultmidpunct}
{\mcitedefaultendpunct}{\mcitedefaultseppunct}\relax
\EndOfBibitem
\bibitem[Chakrabarti \emph{et~al.}(2017)Chakrabarti, Lastras-Monta{\~n}o, Adam,
  Prezioso, Hoskins, Payvand, Madhavan, Ghofrani, Theogarajan,
  Cheng,\emph{et~al.}]{Chakrabarti_etal17}
B.~Chakrabarti, M.~A. Lastras-Monta{\~n}o, G.~Adam, M.~Prezioso, B.~Hoskins,
  M.~Payvand, A.~Madhavan, A.~Ghofrani, L.~Theogarajan, K.-T. Cheng
  \emph{et~al.}, \emph{Scientific Reports}, 2017, \textbf{7}, 42429\relax
\mciteBstWouldAddEndPuncttrue
\mciteSetBstMidEndSepPunct{\mcitedefaultmidpunct}
{\mcitedefaultendpunct}{\mcitedefaultseppunct}\relax
\EndOfBibitem
\bibitem[Jo \emph{et~al.}(2010)Jo, Chang, Ebong, Bhadviya, Mazumder, and
  Lu]{Jo_etal10}
S.~H. Jo, T.~Chang, I.~Ebong, B.~B. Bhadviya, P.~Mazumder and W.~Lu, \emph{Nano
  letters}, 2010, \textbf{10}, 1297--1301\relax
\mciteBstWouldAddEndPuncttrue
\mciteSetBstMidEndSepPunct{\mcitedefaultmidpunct}
{\mcitedefaultendpunct}{\mcitedefaultseppunct}\relax
\EndOfBibitem
\bibitem[Ielmini and Waser(2015)]{Ielmini_Waser15}
D.~Ielmini and R.~Waser, \emph{Resistive Switching: From Fundamentals of
  Nanoionic Redox Processes to Memristive Device Applications}, John Wiley \&
  Sons, 2015\relax
\mciteBstWouldAddEndPuncttrue
\mciteSetBstMidEndSepPunct{\mcitedefaultmidpunct}
{\mcitedefaultendpunct}{\mcitedefaultseppunct}\relax
\EndOfBibitem
\bibitem[Tuma \emph{et~al.}(2016)Tuma, Pantazi, Le~Gallo, Sebastian, and
  Eleftheriou]{Tuma_etal16}
T.~Tuma, A.~Pantazi, M.~Le~Gallo, A.~Sebastian and E.~Eleftheriou, \emph{Nature
  nanotechnology}, 2016, \textbf{11}, 693--699\relax
\mciteBstWouldAddEndPuncttrue
\mciteSetBstMidEndSepPunct{\mcitedefaultmidpunct}
{\mcitedefaultendpunct}{\mcitedefaultseppunct}\relax
\EndOfBibitem
\bibitem[Suri \emph{et~al.}(2013)Suri, Querlioz, Bichler, Palma, Vianello,
  Vuillaume, Gamrat, and DeSalvo]{Suri_etal13}
M.~Suri, D.~Querlioz, O.~Bichler, G.~Palma, E.~Vianello, D.~Vuillaume,
  C.~Gamrat and B.~DeSalvo, \emph{{IEEE} Transactions on Electron Devices},
  2013, \textbf{60}, 2402--2409\relax
\mciteBstWouldAddEndPuncttrue
\mciteSetBstMidEndSepPunct{\mcitedefaultmidpunct}
{\mcitedefaultendpunct}{\mcitedefaultseppunct}\relax
\EndOfBibitem
\bibitem[Suri \emph{et~al.}(2012)Suri, Bichler, Querlioz, Palma, Vianello,
  Vuillaume, Gamrat, and DeSalvo]{Suri_etal12}
M.~Suri, O.~Bichler, D.~Querlioz, G.~Palma, E.~Vianello, D.~Vuillaume,
  C.~Gamrat and B.~DeSalvo, Electron Devices Meeting ({IEDM}), 2012 {IEEE}
  International, 2012, pp. 3--10\relax
\mciteBstWouldAddEndPuncttrue
\mciteSetBstMidEndSepPunct{\mcitedefaultmidpunct}
{\mcitedefaultendpunct}{\mcitedefaultseppunct}\relax
\EndOfBibitem
\bibitem[Gaba \emph{et~al.}(2013)Gaba, Sheridan, Zhou, Choi, and
  Lu]{Gaba_etal13}
S.~Gaba, P.~Sheridan, J.~Zhou, S.~Choi and W.~Lu, \emph{Nanoscale}, 2013,
  \textbf{5}, 5872--5878\relax
\mciteBstWouldAddEndPuncttrue
\mciteSetBstMidEndSepPunct{\mcitedefaultmidpunct}
{\mcitedefaultendpunct}{\mcitedefaultseppunct}\relax
\EndOfBibitem
\bibitem[Jo \emph{et~al.}(2008)Jo, Kim, and Lu]{Jo_etal08}
S.~H. Jo, K.-H. Kim and W.~Lu, \emph{Nano letters}, 2008, \textbf{9},
  496--500\relax
\mciteBstWouldAddEndPuncttrue
\mciteSetBstMidEndSepPunct{\mcitedefaultmidpunct}
{\mcitedefaultendpunct}{\mcitedefaultseppunct}\relax
\EndOfBibitem
\bibitem[Ambrogio \emph{et~al.}(2016)Ambrogio, Balatti, Milo, Carboni, Wang,
  Calderoni, Ramaswamy, and Ielmini]{Ambrogio_etal16a}
S.~Ambrogio, S.~Balatti, V.~Milo, R.~Carboni, Z.-Q. Wang, A.~Calderoni,
  N.~Ramaswamy and D.~Ielmini, \emph{IEEE Transactions on Electron Devices},
  2016, \textbf{63}, 1508--1515\relax
\mciteBstWouldAddEndPuncttrue
\mciteSetBstMidEndSepPunct{\mcitedefaultmidpunct}
{\mcitedefaultendpunct}{\mcitedefaultseppunct}\relax
\EndOfBibitem
\bibitem[Yang \emph{et~al.}(2013)Yang, Strukov, and Stewart]{Yang_etal13}
J.~J. Yang, D.~B. Strukov and D.~R. Stewart, \emph{Nature nanotechnology},
  2013, \textbf{8}, 13--24\relax
\mciteBstWouldAddEndPuncttrue
\mciteSetBstMidEndSepPunct{\mcitedefaultmidpunct}
{\mcitedefaultendpunct}{\mcitedefaultseppunct}\relax
\EndOfBibitem
\bibitem[Truong \emph{et~al.}(2014)Truong, Ham, and Min]{Truong_etal14}
S.~N. Truong, S.-J. Ham and K.-S. Min, \emph{Nanoscale Research Letters}, 2014,
  \textbf{9}, 629\relax
\mciteBstWouldAddEndPuncttrue
\mciteSetBstMidEndSepPunct{\mcitedefaultmidpunct}
{\mcitedefaultendpunct}{\mcitedefaultseppunct}\relax
\EndOfBibitem
\bibitem[Serb \emph{et~al.}(2016)Serb, Bill, Khiat, Berdan, Legenstein, and
  Prodromakis]{Serb_etal16}
A.~Serb, J.~Bill, A.~Khiat, R.~Berdan, R.~Legenstein and T.~Prodromakis,
  \emph{Nature communications}, 2016, \textbf{7}, 12611\relax
\mciteBstWouldAddEndPuncttrue
\mciteSetBstMidEndSepPunct{\mcitedefaultmidpunct}
{\mcitedefaultendpunct}{\mcitedefaultseppunct}\relax
\EndOfBibitem
\bibitem[Nair and Indiveri(2017)]{Nair_Indiveri17}
M.~V. Nair and G.~Indiveri, \emph{A differential memristive current-mode
  circuit}, European patent application EP 17183461.7, 2017, Filed
  27.07.2017\relax
\mciteBstWouldAddEndPuncttrue
\mciteSetBstMidEndSepPunct{\mcitedefaultmidpunct}
{\mcitedefaultendpunct}{\mcitedefaultseppunct}\relax
\EndOfBibitem
\bibitem[Serb \emph{et~al.}(2016)Serb, Redman-White, Papavassiliou, and
  Prodromakis]{Serb_etal16b}
A.~Serb, W.~Redman-White, C.~Papavassiliou and T.~Prodromakis, \emph{IEEE
  Transactions on Circuits and Systems I: Regular Papers}, 2016, \textbf{63},
  827--835\relax
\mciteBstWouldAddEndPuncttrue
\mciteSetBstMidEndSepPunct{\mcitedefaultmidpunct}
{\mcitedefaultendpunct}{\mcitedefaultseppunct}\relax
\EndOfBibitem
\bibitem[Vincent \emph{et~al.}(2014)Vincent, Larroque, Zhao, Romdhane, Bichler,
  Gamrat, Klein, Galdin-Retailleau, and Querlioz]{Vincent_etal14}
A.~Vincent, J.~Larroque, W.~Zhao, N.~B. Romdhane, O.~Bichler, C.~Gamrat, J.-O.
  Klein, S.~Galdin-Retailleau and D.~Querlioz, International Symposium on
  Circuits and Systems, ({ISCAS}), 2014, 2014, pp. 1074--1077\relax
\mciteBstWouldAddEndPuncttrue
\mciteSetBstMidEndSepPunct{\mcitedefaultmidpunct}
{\mcitedefaultendpunct}{\mcitedefaultseppunct}\relax
\EndOfBibitem
\bibitem[Al-Shedivat \emph{et~al.}(2015)Al-Shedivat, Naous, Cauwenberghs, and
  Salama]{Al-Shedivat_etal15a}
M.~Al-Shedivat, R.~Naous, G.~Cauwenberghs and K.~N. Salama, \emph{IEEE Journal
  on Emerging and Selected Topics in Circuits and Systems}, 2015, \textbf{5},
  242--253\relax
\mciteBstWouldAddEndPuncttrue
\mciteSetBstMidEndSepPunct{\mcitedefaultmidpunct}
{\mcitedefaultendpunct}{\mcitedefaultseppunct}\relax
\EndOfBibitem
\bibitem[Neftci \emph{et~al.}(2016)Neftci, Pedroni, Joshi, Al-Shedivat, and
  Cauwenberghs]{Neftci_etal16}
E.~O. Neftci, B.~U. Pedroni, S.~Joshi, M.~Al-Shedivat and G.~Cauwenberghs,
  \emph{Frontiers in Neuroscience}, 2016, \textbf{10}, 241\relax
\mciteBstWouldAddEndPuncttrue
\mciteSetBstMidEndSepPunct{\mcitedefaultmidpunct}
{\mcitedefaultendpunct}{\mcitedefaultseppunct}\relax
\EndOfBibitem
\bibitem[Payvand \emph{et~al.}(2018)Payvand, Muller, and
  Indiveri]{Payvand_etal18}
M.~Payvand, L.~K. Muller and G.~Indiveri, Circuits and Systems (ISCAS), 2018
  IEEE International Symposium on, 2018, pp. 1--5\relax
\mciteBstWouldAddEndPuncttrue
\mciteSetBstMidEndSepPunct{\mcitedefaultmidpunct}
{\mcitedefaultendpunct}{\mcitedefaultseppunct}\relax
\EndOfBibitem
\bibitem[Bill and Legenstein(2014)]{Bill_Legenstein14}
J.~Bill and R.~Legenstein, \emph{Frontiers in neuroscience}, 2014, \textbf{8},
  1--18\relax
\mciteBstWouldAddEndPuncttrue
\mciteSetBstMidEndSepPunct{\mcitedefaultmidpunct}
{\mcitedefaultendpunct}{\mcitedefaultseppunct}\relax
\EndOfBibitem
\bibitem[Courbariaux \emph{et~al.}(2015)Courbariaux, Bengio, and
  David]{Courbariaux_etal15}
M.~Courbariaux, Y.~Bengio and J.-P. David, Advances in neural information
  processing systems, 2015, pp. 3123--3131\relax
\mciteBstWouldAddEndPuncttrue
\mciteSetBstMidEndSepPunct{\mcitedefaultmidpunct}
{\mcitedefaultendpunct}{\mcitedefaultseppunct}\relax
\EndOfBibitem
\bibitem[Muller and Indiveri(2015)]{Muller_Indiveri15}
L.~K. Muller and G.~Indiveri, \emph{arXiv preprint arXiv:1504.05767}, 2015,
  1--11\relax
\mciteBstWouldAddEndPuncttrue
\mciteSetBstMidEndSepPunct{\mcitedefaultmidpunct}
{\mcitedefaultendpunct}{\mcitedefaultseppunct}\relax
\EndOfBibitem
\bibitem[Wozniak \emph{et~al.}(2017)Wozniak, Pantazi, Sidler, Papandreou,
  Leblebici, and Eleftheriou]{Wozniak_etal17}
S.~Wozniak, A.~Pantazi, S.~Sidler, N.~Papandreou, Y.~Leblebici and
  E.~Eleftheriou, \emph{IEEE Transactions on Circuits and Systems II: Express
  Briefs}, 2017,  1342--1346\relax
\mciteBstWouldAddEndPuncttrue
\mciteSetBstMidEndSepPunct{\mcitedefaultmidpunct}
{\mcitedefaultendpunct}{\mcitedefaultseppunct}\relax
\EndOfBibitem
\bibitem[Covi \emph{et~al.}(2016)Covi, Brivio, Serb, Prodromakis, Fanciulli,
  and Spiga]{Covi_etal16}
E.~Covi, S.~Brivio, A.~Serb, T.~Prodromakis, M.~Fanciulli and S.~Spiga,
  \emph{Frontiers in neuroscience}, 2016, \textbf{10}, 1--13\relax
\mciteBstWouldAddEndPuncttrue
\mciteSetBstMidEndSepPunct{\mcitedefaultmidpunct}
{\mcitedefaultendpunct}{\mcitedefaultseppunct}\relax
\EndOfBibitem
\bibitem[Serrano-Gotarredona and
  Linares-Barranco(2014)]{Serrano-Gotarredona_Linares-Barranco14}
T.~Serrano-Gotarredona and B.~Linares-Barranco, \emph{Memristors and Memristive
  Systems}, Springer, 2014, pp. 353--377\relax
\mciteBstWouldAddEndPuncttrue
\mciteSetBstMidEndSepPunct{\mcitedefaultmidpunct}
{\mcitedefaultendpunct}{\mcitedefaultseppunct}\relax
\EndOfBibitem
\bibitem[Indiveri \emph{et~al.}(2013)Indiveri, Linares-Barranco, Legenstein,
  Deligeorgis, and Prodromakis]{Indiveri_etal13}
G.~Indiveri, B.~Linares-Barranco, R.~Legenstein, G.~Deligeorgis and
  T.~Prodromakis, \emph{Nanotechnology}, 2013, \textbf{24}, 384010\relax
\mciteBstWouldAddEndPuncttrue
\mciteSetBstMidEndSepPunct{\mcitedefaultmidpunct}
{\mcitedefaultendpunct}{\mcitedefaultseppunct}\relax
\EndOfBibitem
\bibitem[Liu \emph{et~al.}(2002)Liu, Kramer, Indiveri, Delbruck, and
  Douglas]{Liu_etal02a}
S.-C. Liu, J.~Kramer, G.~Indiveri, T.~Delbruck and R.~Douglas, \emph{Analog
  {VLSI}:Circuits and Principles}, MIT Press, 2002\relax
\mciteBstWouldAddEndPuncttrue
\mciteSetBstMidEndSepPunct{\mcitedefaultmidpunct}
{\mcitedefaultendpunct}{\mcitedefaultseppunct}\relax
\EndOfBibitem
\bibitem[mnist()]{mnist}
\emph{The {MNIST} database of handwritten digits}, Yann LeCun's web-site, 2012,
  \url{http://yann.lecun.com/exdb/mnist/}\relax
\mciteBstWouldAddEndPuncttrue
\mciteSetBstMidEndSepPunct{\mcitedefaultmidpunct}
{\mcitedefaultendpunct}{\mcitedefaultseppunct}\relax
\EndOfBibitem
\bibitem[Deiss \emph{et~al.}(1998)Deiss, Douglas, and Whatley]{Deiss_etal98}
S.~Deiss, R.~Douglas and A.~Whatley, \emph{Pulsed Neural Networks}, MIT Press,
  1998, ch.~6, pp. 157--78\relax
\mciteBstWouldAddEndPuncttrue
\mciteSetBstMidEndSepPunct{\mcitedefaultmidpunct}
{\mcitedefaultendpunct}{\mcitedefaultseppunct}\relax
\EndOfBibitem
\bibitem[Lazzaro and Wawrzynek(1995)]{Lazzaro_Wawrzynek95}
J.~Lazzaro and J.~Wawrzynek, Sixteenth Conference on Advanced Research in
  {VLSI}, 1995, pp. 158--169\relax
\mciteBstWouldAddEndPuncttrue
\mciteSetBstMidEndSepPunct{\mcitedefaultmidpunct}
{\mcitedefaultendpunct}{\mcitedefaultseppunct}\relax
\EndOfBibitem
\bibitem[Boahen(1998)]{Boahen98}
K.~Boahen, \emph{Neuromorphic Systems Engineering}, Kluwer Academic, Norwell,
  MA, 1998, pp. 229--259\relax
\mciteBstWouldAddEndPuncttrue
\mciteSetBstMidEndSepPunct{\mcitedefaultmidpunct}
{\mcitedefaultendpunct}{\mcitedefaultseppunct}\relax
\EndOfBibitem
\bibitem[Nair \emph{et~al.}(2017)Nair, Mueller, and Indiveri]{Nair_etal17}
M.~V. Nair, L.~K. Mueller and G.~Indiveri, \emph{Nano Futures}, 2017,
  \textbf{1}, 1--12\relax
\mciteBstWouldAddEndPuncttrue
\mciteSetBstMidEndSepPunct{\mcitedefaultmidpunct}
{\mcitedefaultendpunct}{\mcitedefaultseppunct}\relax
\EndOfBibitem
\bibitem[Gilbert(1996)]{Gilbert96}
B.~Gilbert, \emph{Analog Integrated Circuits and Signal Processing}, 1996,
  \textbf{9}, 95--118\relax
\mciteBstWouldAddEndPuncttrue
\mciteSetBstMidEndSepPunct{\mcitedefaultmidpunct}
{\mcitedefaultendpunct}{\mcitedefaultseppunct}\relax
\EndOfBibitem
\bibitem[Bartolozzi and Indiveri(2007)]{Bartolozzi_Indiveri07a}
C.~Bartolozzi and G.~Indiveri, \emph{Neural Computation}, 2007, \textbf{19},
  2581--2603\relax
\mciteBstWouldAddEndPuncttrue
\mciteSetBstMidEndSepPunct{\mcitedefaultmidpunct}
{\mcitedefaultendpunct}{\mcitedefaultseppunct}\relax
\EndOfBibitem
\bibitem[Widrow and Hoff(1960)]{Widrow_Hoff60}
B.~Widrow and M.~Hoff, 1960 {IRE} {WESCON} Convention Record, Part 4, New York,
  1960, pp. 96--104\relax
\mciteBstWouldAddEndPuncttrue
\mciteSetBstMidEndSepPunct{\mcitedefaultmidpunct}
{\mcitedefaultendpunct}{\mcitedefaultseppunct}\relax
\EndOfBibitem
\bibitem[Gilbert(1990)]{Gilbert90}
B.~Gilbert, \emph{Analogue {IC} design: the current-mode approach}, Peregrinus,
  Stevenage, Herts., UK, 1990, ch.~2, pp. 11--91\relax
\mciteBstWouldAddEndPuncttrue
\mciteSetBstMidEndSepPunct{\mcitedefaultmidpunct}
{\mcitedefaultendpunct}{\mcitedefaultseppunct}\relax
\EndOfBibitem
\bibitem[Brivio \emph{et~al.}(2016)Brivio, Covi, Serb, Prodromakis, Fanciulli,
  and Spiga]{Brivio_etal16}
S.~Brivio, E.~Covi, A.~Serb, T.~Prodromakis, M.~Fanciulli and S.~Spiga,
  \emph{Applied Physics Letters}, 2016, \textbf{109}, 133504\relax
\mciteBstWouldAddEndPuncttrue
\mciteSetBstMidEndSepPunct{\mcitedefaultmidpunct}
{\mcitedefaultendpunct}{\mcitedefaultseppunct}\relax
\EndOfBibitem
\bibitem[Naous \emph{et~al.}(2016)Naous, Al-Shedivat, and Salama]{Naous_etal16}
R.~Naous, M.~Al-Shedivat and K.~N. Salama, \emph{IEEE Transactions on
  Nanotechnology}, 2016, \textbf{15}, 15--28\relax
\mciteBstWouldAddEndPuncttrue
\mciteSetBstMidEndSepPunct{\mcitedefaultmidpunct}
{\mcitedefaultendpunct}{\mcitedefaultseppunct}\relax
\EndOfBibitem
\bibitem[Hertz \emph{et~al.}(1991)Hertz, Krogh, and Palmer]{Hertz_etal91}
J.~Hertz, A.~Krogh and R.~Palmer, \emph{Introduction to the Theory of Neural
  Computation}, Addison-Wesley, Reading, MA, 1991\relax
\mciteBstWouldAddEndPuncttrue
\mciteSetBstMidEndSepPunct{\mcitedefaultmidpunct}
{\mcitedefaultendpunct}{\mcitedefaultseppunct}\relax
\EndOfBibitem
\bibitem[LeCun \emph{et~al.}(2015)LeCun, Bengio, and Hinton]{LeCun_etal15}
Y.~LeCun, Y.~Bengio and G.~Hinton, \emph{Nature}, 2015, \textbf{521},
  436--444\relax
\mciteBstWouldAddEndPuncttrue
\mciteSetBstMidEndSepPunct{\mcitedefaultmidpunct}
{\mcitedefaultendpunct}{\mcitedefaultseppunct}\relax
\EndOfBibitem
\bibitem[Schmidhuber(2015)]{Schmidhuber15}
J.~Schmidhuber, \emph{Neural Networks}, 2015, \textbf{61}, 85--117\relax
\mciteBstWouldAddEndPuncttrue
\mciteSetBstMidEndSepPunct{\mcitedefaultmidpunct}
{\mcitedefaultendpunct}{\mcitedefaultseppunct}\relax
\EndOfBibitem
\bibitem[Schemmel \emph{et~al.}(2010)Schemmel, Bruderle, Grubl, Hock, Meier,
  and Millner]{Schemmel_etal10}
J.~Schemmel, D.~Bruderle, A.~Grubl, M.~Hock, K.~Meier and S.~Millner, Circuits
  and Systems (ISCAS), Proceedings of 2010 IEEE International Symposium on,
  2010, pp. 1947--1950\relax
\mciteBstWouldAddEndPuncttrue
\mciteSetBstMidEndSepPunct{\mcitedefaultmidpunct}
{\mcitedefaultendpunct}{\mcitedefaultseppunct}\relax
\EndOfBibitem
\bibitem[Raghavan and Tompson(1987)]{Raghavan_Tompson87}
P.~Raghavan and C.~D. Tompson, \emph{Combinatorica}, 1987, \textbf{7},
  365--374\relax
\mciteBstWouldAddEndPuncttrue
\mciteSetBstMidEndSepPunct{\mcitedefaultmidpunct}
{\mcitedefaultendpunct}{\mcitedefaultseppunct}\relax
\EndOfBibitem
\bibitem[Brader \emph{et~al.}(2007)Brader, Senn, and Fusi]{Brader_etal07}
J.~M. Brader, W.~Senn and S.~Fusi, \emph{Neural computation}, 2007,
  \textbf{19}, 2881--2912\relax
\mciteBstWouldAddEndPuncttrue
\mciteSetBstMidEndSepPunct{\mcitedefaultmidpunct}
{\mcitedefaultendpunct}{\mcitedefaultseppunct}\relax
\EndOfBibitem
\bibitem[Sheik \emph{et~al.}(2017)Sheik, Paul, Augustine, and
  Cauwenberghs]{Sheik_etal17}
S.~Sheik, S.~Paul, C.~Augustine and G.~Cauwenberghs, \emph{arXiv preprint
  arXiv:1701.01495}, 2017\relax
\mciteBstWouldAddEndPuncttrue
\mciteSetBstMidEndSepPunct{\mcitedefaultmidpunct}
{\mcitedefaultendpunct}{\mcitedefaultseppunct}\relax
\EndOfBibitem
\bibitem[Baldassi \emph{et~al.}(2016)Baldassi, Gerace, Lucibello, Saglietti,
  and Zecchina]{Baldassi_etal16}
C.~Baldassi, F.~Gerace, C.~Lucibello, L.~Saglietti and R.~Zecchina, \emph{Phys.
  Rev. E}, 2016, \textbf{93}, 052313\relax
\mciteBstWouldAddEndPuncttrue
\mciteSetBstMidEndSepPunct{\mcitedefaultmidpunct}
{\mcitedefaultendpunct}{\mcitedefaultseppunct}\relax
\EndOfBibitem
\bibitem[Goodman and Brette(2009)]{Goodman_Brette09}
D.~Goodman and R.~Brette, \emph{Frontiers in Neuroscience}, 2009, \textbf{3},
  192--197\relax
\mciteBstWouldAddEndPuncttrue
\mciteSetBstMidEndSepPunct{\mcitedefaultmidpunct}
{\mcitedefaultendpunct}{\mcitedefaultseppunct}\relax
\EndOfBibitem
\bibitem[Bishop(2006)]{Bishop06}
C.~Bishop, \emph{{Pattern recognition and machine learning}}, Springer New
  York, 2006\relax
\mciteBstWouldAddEndPuncttrue
\mciteSetBstMidEndSepPunct{\mcitedefaultmidpunct}
{\mcitedefaultendpunct}{\mcitedefaultseppunct}\relax
\EndOfBibitem
\bibitem[Stathopoulos \emph{et~al.}(2017)Stathopoulos, Khiat, Trapatseli,
  Cortese, Serb, Valov, and Prodromakis]{Stathopoulos_etal17}
S.~Stathopoulos, A.~Khiat, M.~Trapatseli, S.~Cortese, A.~Serb, I.~Valov and
  T.~Prodromakis, \emph{Scientific reports}, 2017, \textbf{7}, 17532\relax
\mciteBstWouldAddEndPuncttrue
\mciteSetBstMidEndSepPunct{\mcitedefaultmidpunct}
{\mcitedefaultendpunct}{\mcitedefaultseppunct}\relax
\EndOfBibitem
\bibitem[Frascaroli \emph{et~al.}(2018)Frascaroli, Brivio, Covi, and
  Spiga]{frascaroli_etal18}
J.~Frascaroli, S.~Brivio, E.~Covi and S.~Spiga, \emph{Scientific reports},
  2018, \textbf{8}, 71--78\relax
\mciteBstWouldAddEndPuncttrue
\mciteSetBstMidEndSepPunct{\mcitedefaultmidpunct}
{\mcitedefaultendpunct}{\mcitedefaultseppunct}\relax
\EndOfBibitem
\bibitem[Chang \emph{et~al.}(2011)Chang, Jo, Kim, Sheridan, Gaba, and
  Lu]{chang_etal11}
T.~Chang, S.-H. Jo, K.-H. Kim, P.~Sheridan, S.~Gaba and W.~Lu, \emph{Applied
  physics A}, 2011, \textbf{102}, 857--863\relax
\mciteBstWouldAddEndPuncttrue
\mciteSetBstMidEndSepPunct{\mcitedefaultmidpunct}
{\mcitedefaultendpunct}{\mcitedefaultseppunct}\relax
\EndOfBibitem
\bibitem[Prezioso \emph{et~al.}(2015)Prezioso, Kataeva, Merrikh-Bayat, Hoskins,
  Adam, Sota, Likharev, and Strukov]{prezioso_etal15b}
M.~Prezioso, I.~Kataeva, F.~Merrikh-Bayat, B.~Hoskins, G.~Adam, T.~Sota,
  K.~Likharev and D.~Strukov, IEEE International Electron Devices Meeting
  (IEDM), 2015, pp. 209--223\relax
\mciteBstWouldAddEndPuncttrue
\mciteSetBstMidEndSepPunct{\mcitedefaultmidpunct}
{\mcitedefaultendpunct}{\mcitedefaultseppunct}\relax
\EndOfBibitem
\bibitem[Breiman(1996)]{Breiman_96bagging}
L.~Breiman, \emph{Machine learning}, 1996, \textbf{24}, 123--140\relax
\mciteBstWouldAddEndPuncttrue
\mciteSetBstMidEndSepPunct{\mcitedefaultmidpunct}
{\mcitedefaultendpunct}{\mcitedefaultseppunct}\relax
\EndOfBibitem
\bibitem[Schapire(1990)]{Schapire_90boosting}
R.~E. Schapire, \emph{Machine learning}, 1990, \textbf{5}, 197--227\relax
\mciteBstWouldAddEndPuncttrue
\mciteSetBstMidEndSepPunct{\mcitedefaultmidpunct}
{\mcitedefaultendpunct}{\mcitedefaultseppunct}\relax
\EndOfBibitem
\bibitem[Freund and Schapire(1997)]{Freund_etal97adaboost}
Y.~Freund and R.~E. Schapire, \emph{Journal of computer and system sciences},
  1997, \textbf{55}, 119--139\relax
\mciteBstWouldAddEndPuncttrue
\mciteSetBstMidEndSepPunct{\mcitedefaultmidpunct}
{\mcitedefaultendpunct}{\mcitedefaultseppunct}\relax
\EndOfBibitem
\bibitem[Pi \emph{et~al.}(2013)Pi, Lin, and Xia]{pi_etal13cross}
S.~Pi, P.~Lin and Q.~Xia, \emph{Journal of Vacuum Science \& Technology B,
  Nanotechnology and Microelectronics: Materials, Processing, Measurement, and
  Phenomena}, 2013, \textbf{31}, 06FA02\relax
\mciteBstWouldAddEndPuncttrue
\mciteSetBstMidEndSepPunct{\mcitedefaultmidpunct}
{\mcitedefaultendpunct}{\mcitedefaultseppunct}\relax
\EndOfBibitem
\bibitem[Govoreanu \emph{et~al.}(2013)Govoreanu, Redolfi, Zhang, Adelmann,
  Popovici, Clima, Hody, Paraschiv, Radu,
  Franquet,\emph{et~al.}]{govoreanu_etal13vacancy}
B.~Govoreanu, A.~Redolfi, L.~Zhang, C.~Adelmann, M.~Popovici, S.~Clima,
  H.~Hody, V.~Paraschiv, I.~Radu, A.~Franquet \emph{et~al.}, Electron Devices
  Meeting (IEDM), 2013 IEEE International, 2013, pp. 10--2\relax
\mciteBstWouldAddEndPuncttrue
\mciteSetBstMidEndSepPunct{\mcitedefaultmidpunct}
{\mcitedefaultendpunct}{\mcitedefaultseppunct}\relax
\EndOfBibitem
\bibitem[Li \emph{et~al.}(2018)Li, Hu, Li, Jiang, Ge, Montgomery, Zhang, Song,
  D{\'a}vila, Graves,\emph{et~al.}]{li_etal18analogue}
C.~Li, M.~Hu, Y.~Li, H.~Jiang, N.~Ge, E.~Montgomery, J.~Zhang, W.~Song,
  N.~D{\'a}vila, C.~E. Graves \emph{et~al.}, \emph{Nature Electronics}, 2018,
  \textbf{1}, 52--59\relax
\mciteBstWouldAddEndPuncttrue
\mciteSetBstMidEndSepPunct{\mcitedefaultmidpunct}
{\mcitedefaultendpunct}{\mcitedefaultseppunct}\relax
\EndOfBibitem
\bibitem[Moradi \emph{et~al.}(2015)Moradi, Indiveri, Qiao, and
  Stefanini]{Moradi_etal15}
S.~Moradi, G.~Indiveri, N.~Qiao and F.~Stefanini, \emph{Networks and
  hierarchical routing fabrics with heterogeneous memory structures for
  scalable event-driven computing systems}, European patent application EP
  15/165272, 2015, Filed 27.04.2015\relax
\mciteBstWouldAddEndPuncttrue
\mciteSetBstMidEndSepPunct{\mcitedefaultmidpunct}
{\mcitedefaultendpunct}{\mcitedefaultseppunct}\relax
\EndOfBibitem
\bibitem[Fasnacht and Indiveri(2011)]{Fasnacht_Indiveri11}
D.~Fasnacht and G.~Indiveri, Conference on Information Sciences and Systems,
  {CISS} 2011, Johns Hopkins University, 2011, pp. 1--6\relax
\mciteBstWouldAddEndPuncttrue
\mciteSetBstMidEndSepPunct{\mcitedefaultmidpunct}
{\mcitedefaultendpunct}{\mcitedefaultseppunct}\relax
\EndOfBibitem
\bibitem[Serrano-Gotarredona \emph{et~al.}(2009)Serrano-Gotarredona, Oster,
  Lichtsteiner, Linares-Barranco, Paz-Vicente, G{\'o}mez-Rodriguez,
  Camunas-Mesa, Berner, Rivas-Perez, Delbruck, Liu, Douglas, H{\"a}fliger,
  Jimenez-Moreno, Civit-Ballcels, Serrano-Gotarredona, Acosta-Jim{\'e}nez, and
  Linares-Barranco]{Serrano-Gotarredona_etal09}
R.~Serrano-Gotarredona, M.~Oster, P.~Lichtsteiner, A.~Linares-Barranco,
  R.~Paz-Vicente, F.~G{\'o}mez-Rodriguez, L.~Camunas-Mesa, R.~Berner,
  M.~Rivas-Perez, T.~Delbruck, S.-C. Liu, R.~Douglas, P.~H{\"a}fliger,
  G.~Jimenez-Moreno, A.~Civit-Ballcels, T.~Serrano-Gotarredona,
  A.~Acosta-Jim{\'e}nez and B.~Linares-Barranco, \emph{{IEEE} Transactions on
  Neural Networks}, 2009, \textbf{20}, 1417--1438\relax
\mciteBstWouldAddEndPuncttrue
\mciteSetBstMidEndSepPunct{\mcitedefaultmidpunct}
{\mcitedefaultendpunct}{\mcitedefaultseppunct}\relax
\EndOfBibitem
\bibitem[Park \emph{et~al.}(2016)Park, Yu, Joshi, Maier, and
  Cauwenberghs]{Park_etal16}
J.~Park, T.~Yu, S.~Joshi, C.~Maier and G.~Cauwenberghs, \emph{{IEEE}
  Transactions on Neural Networks and Learning Systems}, 2016,  1--15\relax
\mciteBstWouldAddEndPuncttrue
\mciteSetBstMidEndSepPunct{\mcitedefaultmidpunct}
{\mcitedefaultendpunct}{\mcitedefaultseppunct}\relax
\EndOfBibitem
\end{mcitethebibliography}
\bibliographystyle{faraday18}

\end{document}